\documentclass[%
 reprint,
 amsmath,amssymb,
 aps,
]{revtex4-1}

\usepackage{multirow}
\usepackage{graphicx}
\usepackage{dcolumn}
\usepackage{bm}
\usepackage{natbib}
\usepackage{url}
\usepackage{subcaption}
\usepackage[colorlinks]{hyperref}
\begin{document}

\preprint{APS/123-QED}

\title{Models of Continuous-Time Networks with Tie Decay, Diffusion, and Convection}

\author{Xinzhe Zuo$^{1}$ and Mason A Porter$^1$}
\affiliation{%
 $^1$Department of Mathematics, UCLA, Los Angeles, California 90095, USA
}%

\date{\today}

\begin{abstract}

The study of temporal networks in discrete time has yielded numerous insights into time-dependent networked systems in a wide variety of applications. For many complex systems, however, it is useful to develop continuous-time models of networks and to compare them to associated discrete models. In this paper, we study several continuous-time network models and examine discrete approximations of them both numerically and analytically. To consider continuous-time networks, we associate each edge in a graph with a time-dependent tie strength that can take continuous non-negative values and decays in time after the most recent interaction. We investigate how the mean tie strength evolves with time in several models, and we explore --- both numerically and analytically --- criteria for the emergence of a giant connected component in some of these models. We also briefly examine the effects of interaction patterns of our continuous-time networks on contagion dynamics in a susceptible--infected--recovered model of an infectious disease.

\end{abstract}

\maketitle




\section{Introduction}
\label{sec:level1}

Networks, in the form of graphs or more complicated structures, are useful models of many complex systems in nature, society, and technology \cite{newman2018,vision2020}. In the simplest case of a time-independent graph, one models entities as nodes and interactions between them as (possibly weighted and/or directed) edges. However, most networks change in time, and the study of so-called ``temporal networks'' --- in which nodes and/or edges change in time --- is one of the most active areas of network science \cite{holme2012temporal,holme2015,Holme2019}.

Temporal networks differ from time-independent networks in several respects. One significant feature is that the edges of a temporal network may change between active and inactive states. For example, in a communication network, e-mails or text messages may yield instantaneous interactions between pairs of entities, where we consider an edge to be active during instantaneous communication. In other situations, such as in a phone call, interactions between entities of a social network may be active for some finite duration of time. Temporal networks are very popular for studying time-dependent networked systems, but almost all formulations of them have focused on discrete time \cite{ahmad2018tie}. However, it is more appropriate to study many systems using continuous-time temporal networks, which allow both discrete and continuous ties. Indeed, even when interactions are instantaneous, their importance or influence may last beyond the interaction time itself, and one can model them as decaying continuously as a function of time \cite{burt2000decay,lerman2016information,lerman2012social}. In such a ``tie-decay network'' framework, as advocated in \cite{ahmad2018tie}, one separates the concepts of interactions and ties between entities. An interaction may or may not be instantaneous (depending on the model), but the existence and weights of the ties between entities --- which are affected by the interactions --- change continuously in time.

Ties between entities of a social network strengthen with repeated interactions, and they often deteriorate without such interactions~\cite{burt2000decay,moro2017}. Our work is motivated by the recent formalization of tie-decay networks by Ahmad et al. \cite{ahmad2018tie}. In this study, the strength of a tie between nodes decays exponentially in the absence of interactions, and discrete interactions between entities boost the strength of a tie between entities. This mechanism is also reminiscent of models of Hebbian learning in neuronal networks, as the tie strength between neurons can increase when they have similar interaction patterns \cite{hebbian2011}. In the context of social networks, Jin et al. \cite{jin2001} examined continuous-time networks with an exponential decay of tie strengths that they used to represent friendship strengths between people in a social network. As we discuss in the present paper, there are various ways to formulate models of tie-decay networks, and we consider a few of them. Another approach for studying temporal networks in both continuous and discrete time is through statistical models, such as exponential random-graph models \cite{fritz2019}.

As discussed in \cite{ahmad2018tie}, a key challenge of studying continuous-time temporal networks is the aggregation of interactions between entities into time windows. There is a delicate balance between smoothing noise and preserving information content, and the choice of the size of a time window plays an important role. If a time window is too small, one may be unable to capture some important features of a network. However, if the time window is too large, it may eclipse important interactions in a network. Given these issues, Sulo et al. illustrated that it is important to examine multiple resolutions in time-dependent networks \cite{sulo2010meaningful}. In the present paper, we focus on the decay and boosting behavior of ties between pairs of nodes. Therefore, it is often more meaningful to examine the time step and the decay rate together, instead of studying them separately. 

To improve the understanding of tie-decay networks, it is important to generalize well-known network models to this setting. A key example is Erd\H{o}s--R\'{e}nyi (ER) networks \cite{erdosrenyi1960evolution,newman2018}, the simplest type of random graph. Each edge in an $G(n,p)$ ER graph exists with a homogeneous, independent probability $p$. An important feature of the $G(n,p)$ model is the emergence of a giant connected component (GCC), which scales linearly with the number $n$ of nodes in the network, for probabilities above some critical value \cite{erdosrenyi1960evolution, newman2018}. A related idea, which has been used in models of numerous phenomena, is percolation on ER graphs and other networks \cite{saberi2015}. Many scholars have studied GCCs (and giant percolating components) in a diverse set of applications, such as navigability in transportation networks \cite{lopez2007} and transmissibility of diseases in social networks \cite{pastor2015}. Salient to the present paper is the work of \cite{jin2001}, who examined the development of a GCC in a model for the formation of a social network.

In the present paper, we incorporate the $G(n,p)$ model into several continuous-time network models using a variety of different mechanisms for the growth and decay of the tie strengths between nodes. These mechanisms include the tie-decay model of Ahmad el al. \cite{ahmad2018tie} and the back-to-unity model of Jin et al. \cite{jin2001}. We also study two mechanisms --- a diffusion model and a convection--diffusion model --- that are inspired by random walks and partial differential equations (PDEs). For all four of these mechanisms, we assume that the tie strength between a pair of nodes is independent of the tie strengths of any other edges in a network. With this independence assumption, we derive the moments of the tie strength {at stationarity} for these models and then compare these results with numerical simulations. We also study the emergence of a GCC in the back-to-unity model, the diffusion model, and the convection--diffusion model. Our results give insights into several different formulations of tie-decay networks, and we see that their properties can differ from each other in substantive ways. As a case study, we also briefly examine the effects of interaction patterns of the back-to-unity model on contagion dynamics in a susceptible--infected--recovered (SIR) model of an infectious disease.

Our paper proceeds as follows. In Section \ref{Model}, we discuss four models of continuous-time tie-decay networks: the recent model of Ahmad et al. \cite{ahmad2018tie}, the back-to-unity model of Jin et al. \cite{jin2001}, and two novel models. We examine the moments of tie strength in the Ahmad et al. model in the long-time limit. {We also study the emergence of a GCC in a particular limit and compare it with our numerical simulations. We then study the moments of the tie strength and the emergence of a GCC in the back-to-unity model.} We also introduce two tie-decay models based on random walks --- one is a diffusion model and the other is a convection--diffusion model --- and we examine GCCs in them using ideas from PDEs and numerical analysis. In Section \ref{sec:sir}, we examine SIR dynamics on the back-to-unity model. In Section \ref{conclusions}, we summarize our results and suggest several future directions.


\section{Models}\label{Model}


\subsection{Tie-decay model of Ahmad et al. \cite{ahmad2018tie}} \label{tie-decay model}

We start with the tie-decay model of Ahmad et al. \cite{ahmad2018tie}. This model yields a graph $G(n,p,\alpha,T)$ with four parameters: $n$ is the number of nodes, $T$ is the total computation time, $\alpha$ is a decay parameter, and $p$ is the probability for a pair of nodes to interact during one time step. This model makes a point of separating the concepts of ``interactions'' and ``ties'', which traditionally are treated as equivalent concepts. There is an underlying continuous time, which we measure in small increments $\delta t$, and a pair of nodes can interact during a time step. The strength of a tie between a pair of nodes depends on the history of interactions. The primary goal of \cite{ahmad2018tie} was to generalize PageRank centrality \cite{gleich2015} to tie-decay networks and apply it a Twitter network as a case study. In our work on this tie-decay model, we consider networks with undirected edges and tie strengths. We focus on the situation in which nodes in a network have an equal probability of interacting with any of the other nodes in each time step. Using a characteristic function, we derive the moments of the tie strength in the long-time limit. We also examine the criterion for the emergence of a GCC in the network in a particular limit.

\begin{figure}[h!] 
  \centering
  \includegraphics[width=0.5\textwidth]{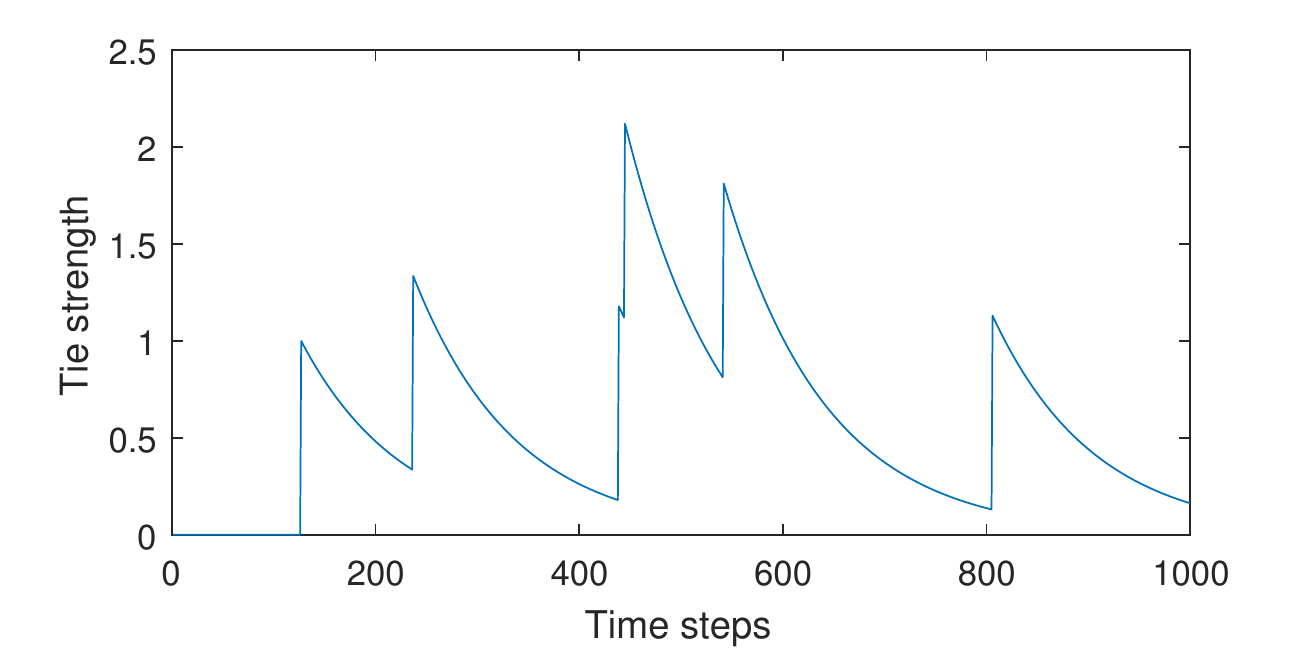}
  \caption{An illustration of dynamics in the tie-decay model of Ahmad et al. \cite{ahmad2018tie} The tie strength between a pair of nodes increases by $1$ when there is an interaction during a time step, and it decays exponentially when there is no interaction. In this simulation, we have $n =1000$ nodes, a decay rate of $\alpha=0.01$, an interaction probability of $p=0.003$, and $T = 1000$ time steps. The vertical axis shows the tie strength of one edge. Six interactions occur between the two nodes that are incident to this edge.}
  \label{fig:typical_tie_decay}
\end{figure}

There are numerous possible choices in the above tie-decay setting, and we follow those of \cite{ahmad2018tie}. During a time step of length $\delta t$, if a pair of nodes interacts, which occurs with a homogeneous probability $p$, the tie strength of the edge between this pair increases by $1$. If they do not interact, which occurs with complementary probability $1 - p$, the strength of the tie between them decays by a factor of $e^{-\alpha \,\delta t}$. We also impose the assumption that, during a single time step, a pair of nodes either has one interaction (thereby increasing the strength of the tie between them) or zero interactions (such that the tie strength between them decays). We suppose that the growth and decay pattern of each pair of nodes is independent of all other pairs, so we independently consider each node pair during each time step. As we mentioned in Section \ref{sec:level1}, it is more appropriate to examine the time step and the decay rate together, rather than separately. For simplicity, we take $\delta t=1$ in this model (and also in the back-to-unity model, which we discuss in \ref{back to unity growth model}). In Figure \ref{fig:typical_tie_decay}, we show an illustrative example of the model's dynamics.

Let $\mathbf{A}$ be an adjacency matrix that is associated with a graph from $G(n,p,\alpha,T)$ and encodes the tie strengths of the edges. The entry $\mathbf{A}_{ef}$ gives the tie strength between nodes $e$ and $f$ (where $e \neq f$). The matrix $\mathbf{A}$ is symmetric and has $0$ entries on the diagonal.

The tie strength of each edge satisfies the same probability distribution, so let us focus on a single edge. Let $s_t$ be the tie strength of a particular edge at time $t$, and suppose that $s_0=0$. To study the model of \cite{ahmad2018tie} with $\delta t = 1$, we run a Monte Carlo simulation for a total of $T$ steps using the following the update rule:
\begin{equation}
	s_{t+1} = \begin{cases}
	s_t + 1\,, \text{   with probability $p$\,,}\\
	s_t  e^{-\alpha}\,, \text{   with probability $(1-p)$}\,. \label{eq:update_rule}
\end{cases}
\end{equation}
That is, 
\begin{equation*}
	s_{t+1} = z_t + e^{-\alpha(1-z_t)}s_{t}\,,
\end{equation*}
where $z_t$ is a Bernoulli random variable with parameter $p$. 

To calculate the expectation of $s_t$, we write
\begin{align*} 
	\mathbb{E}[s_0] &= 0\,, \\
\mathbb{E}[s_1] &= p(1+\mathbb{E}[s_0])+e^{-\alpha}\mathbb{E}[s_0](1-p)\,,\\
&\vdots\\
	\mathbb{E}[s_{t}] &= p(1+\mathbb{E}[s_{t-1}])+e^{-\alpha}\mathbb{E}[s_{t-1}](1-p)\,,\text{   for $t \geq 0$}\,.
\end{align*}
 It is difficult to evaluate the above recursive expression to obtain a closed-form expression for $\mathbb{E}[s_t]$, but we can obtain a good approximation for large times $t$. The expression for $\mathbb{E}[s_t]$ is a sum of terms of the form $p^i e^{-j \alpha}$, where $i \in \{1, \dots, t\}$ and $j \in \{0, \dots, t-1\}$.
 The coefficients of $p^i e^{-j \alpha}$ are all equal to $1$ when $i+j\leq t$, and we can discard the other terms as small as $t\rightarrow \infty$. This allows us to approximate $\mathbb{E}[s_t]$ as follows: 
\begin{align} \label{Expectation}
    \mathbb{E}[s_t] & \approx \sum_{i=1}^t \sum_{j=0}^{n-i}p^i e^{-j \alpha} \nonumber \\
    &= \frac{1}{1-\sigma}\left[\frac{p-p^{t+1}}{1-p}-p\sigma^n \left(\frac{\frac{p}{\sigma}^n - 1}{\frac{p}{\sigma}-1}\right)\right]\,, 
\end{align}
where $\sigma = e^{-\alpha}$. This also yields the long-time behavior of $\mathbb{E}[s_t]$, which is given by 
\begin{equation}
\lim_{t\rightarrow \infty} \mathbb{E}[s_t] = \frac{1}{1-\sigma} \frac{p}{1-p}\,. \label{limit}
\end{equation}
In the long-time limit, which is a stationary state, we can write down the characteristic function of the distribution of {$s := \lim_{t  \to \infty} s_t$}. This function is 
\begin{equation}
    \phi_s(k) = \mathbb{E}[e^{\mathrm{i}ks}]\,,
\end{equation}
where $\mathrm{i}^2 = -1$. Based on our tie-decay interaction and assuming stationary states, it follows that 
\begin{equation}
	\phi_s(k) = p e^{-\mathrm{i}k} \phi_s(k)+(1-p)\phi_s(\sigma k)\,.\label{eq:characteristic_func}
\end{equation}
We do dot possess a closed-form solution to Eq.~\eqref{eq:characteristic_func}. However, we can obtain all of the moments of $s$ by differentiating Eq.~\eqref{eq:characteristic_func} and using the initial condition $\phi_s(0)=1$. The $n^{\text{th}}$ derivative of $\phi_s$ at $k=0$ is 
\begin{equation}
    {\phi_s}^{(n)}(0) = \frac{  p \left( \sum_{j=1}^n {n \choose j} \phi_s^{(n-j)}(0)\mathrm{i}^j \right) }{(1-p)(1-\sigma^n)} \,, \label{eq:moments_tie_decay}
\end{equation}
where $\phi_s^{(j)}(0)$ is the $j^{\text{th}}$ derivative of $\phi_s$ evaluated at $0$. Using Eq.~\eqref{eq:moments_tie_decay}, we calculate the mean $\mathbb{E}[s]$ and variance $\mathrm{var}(s)$ of $s$  to be 
\begin{align*}
    \mathbb{E}[s] &= \frac{p}{(1-\sigma)(1-p)}\,,\\
    \mathrm{var}(s) &= \frac{p}{(1-\sigma^2)(1-p)^2}\,.
\end{align*}
We thereby recover Eq.~\eqref{limit}. 

To verify that we indeed reach a stationary state as $t\rightarrow \infty$, the map with $x \mapsto (p(x+1)+(1-p)\sigma x)$ is a contraction when $\alpha >0$ and $p<1$. Let $x,y\in \mathbb{R}$ and $\phi(x) = (p(x+1)+(1-p)\sigma x)$. It then follows that
\begin{align}
	    \big| \phi(x) -\phi(y) \big| &= \big|p(x-y)+(1-p)\sigma (x-y)\big| \nonumber \\
   & \leq \big| x-y\big| \big|p+(1-p)\sigma \big| \,.\nonumber
\end{align}
By the Banach fixed-point theorem, we achieve a stationary state by iteration.

We now examine how well Eq.~\eqref{Expectation} and Eq.~\eqref{limit} agree using direct numerical simulations of tie-decay networks. We use parameter values of $n=3000$ nodes, a connection probability of $p=0.1$, and a decay rate of $\alpha=0.05$. Equation \eqref{limit} yields a limiting expectation value of $2.2782$. At $t=50$, our numerical computations yield $s_t \approx 2.0368$, and our analytical approximation yields $2.0902$; at $t=100$, we get $s_t \approx 2.2524$, and our analytical approximation yields $2.2628$; at $t=150$, we compute that $s_t \approx 2.2751$, and our analytical approximation yields $2.2770$; at $t=500$, we calculate $s_t \approx 2.2782$, and our analytical approximation yields $2.2782$.

As $t$ becomes larger, the simulations and approximation become progressively closer to each other. We observe that the approximation is always larger than the simulation, because the coefficients, $p^i e^{-j \alpha}$ with $i+j = t+1$, of the largest terms that we dropped in our approximation are always negative. If we include terms of this order (with $i+j=t+1$) in our sum, our refined approximation is smaller than our simulation results, because the coefficients of the next-largest terms ($p^i e^{-j \alpha}$ with $i+j = t+2$) in the sum are always positive. Based on our numerical computations, we observe that these positive and negative corrections to our approximation balance each other, rendering Eq.~\eqref{limit} an accurate approximation in the long-time limit.

As we noted previously, we do not possess a closed form for the characteristic function of the distribution \eqref{eq:characteristic_func} and the distribution of tie strengths. However, we can approximate this tie strength in some limit by formulating the problem into a Poisson process for large $T$ and small $p$. 

Consider a Poisson process with mean and variance $\lambda = Tp$. The tie strength decays exponentially over time until the Poisson process experiences an arrival, with which the tie strength instantaneously increases by $1$. This is an equivalent formulation of the tie-decay process with a total simulation time. The number $N_T$ of arrivals over time $T$ for the Poisson process follows the Poisson distribution 
\begin{equation}
    \mathbf{P}(N_T = j) = \frac{\lambda^j \exp{(-\lambda)}}{j!}\,.
\end{equation}

Let $s$ be the tie strength of an edge at the end (specifically, with $t \rightarrow \infty$) of a tie-decay process that starts at $s_0\geq 0$. By the law of total probability, 
\begin{equation}
    \mathbf{P}(s< \Tilde{s}) = \sum_{j=0}^\infty \mathbf{P}(s<\Tilde{s}|N_T = j) \mathbf{P}(N_T = j)\,.
\end{equation}
The case in which $N_T=0$ is not very interesting, as the tie strength just decays exponentially. When $N_T=1$, let $\tau$ be the (unique) arrival time of the Poisson process. If $\Tilde{s}\geq (s_0+1)\exp(-T\alpha)$ (where equality holds when the arrival occurs at $t=0$), it follows that
 \begin{equation}
    \{s<\Tilde{s}\} \iff \{ \tau < \frac{1}{\alpha}\log(\Tilde{s}\exp(T\alpha)-s_0)\}\,. \label{eq:arrival_time_condition}
\end{equation}
The logical statement~\eqref{eq:arrival_time_condition} suggests that we can readily calculate the distribution of $\tau$, as there is a unique arrival during the interval. We have 
\begin{equation}
    \mathbf{P}(\tau \leq t | N_T =1 ) = \frac{t}{T}\,. \label{eq:one_arrival_time}
\end{equation}
From \eqref{eq:arrival_time_condition} and \eqref{eq:one_arrival_time}, we obtain
\begin{equation}
    \mathbf{P}(s<\Tilde{s}|N_T=1) = \begin{cases}
    \frac{\log(\Tilde{s}e^{T\alpha}-s_0)\}}{\alpha T}\,, \quad &\Tilde{s}\geq (s_0+1)e^{-T\alpha} \\
    0\,, \quad &\text{otherwise}\,.
    \end{cases}
\end{equation}

When $\lambda \ll 1$, we can approximate the tie-decay process by assuming that $\mathbf{P}(N_T\geq 2)=0$. That is, we are assuming that each node forms at most one tie during the entire process. In this scenario, suppose that $\Tilde{s}\geq (s_0+1)\exp(-T\alpha)$. It then follows that
\begin{align}\label{eq:tiedecay_approx}
    \mathbf{P}(s<\Tilde{s}) &\approx \sum_{j=0}^1 \mathbf{P}(s<\Tilde{s}|N_T = j) \mathbf{P}(N_T = j)\nonumber \\
    &= e^{-Tp}\left(1+\frac{p}{\alpha}\log(\Tilde{s}e^{T\alpha}-s_0)\right)\,.
\end{align}

{We now impose a threshold $g$ for the tie strength, such that we only consider edges with tie strengths that are at least $g$ to be active.

Setting $\Tilde{s} = g$, we approximate the value of a critical threshold $g_{\text{crit}}$ for the emergence of a GCC in a tie-decay network. We write
\begin{equation}
    g_{\text{crit}} = \exp{\left(\frac{\alpha}{p}\left(e^{Tp}(1-\frac{1}{n})-1\right)-T\alpha\right)} + s_0 e^{-T\alpha}\,.\label{eq:crit_threshold}
\end{equation}
If $g < g_{\text{crit}}$, there is a GCC in our tie-decay network; if $g > g_{\text{crit}}$, there is not a GCC. In Fig.~\ref{fig:crit_threshold}, we examine the effect of the $\alpha$ on $g_{\text{crit}}$. We calculate that the critical threshold $g_{\text{crit}}$ for a GCC to emerge are $g_{\text{crit}} \approx 0.9502$, $g_{\text{crit}} \approx0.6345$, and $g_{\text{crit}} \approx 0.0106$ for decay rates of $\alpha = 0.001$, $\alpha = 0.01$, and $\alpha = 0.1$, respectively. As we can see with our simulations, we expect to observe a phase transition at $g=g_{\text{crit}}$.

\begin{figure*}
\begin{subfigure}[t]{0.32\textwidth}
        \centering
        \includegraphics[width=\linewidth]{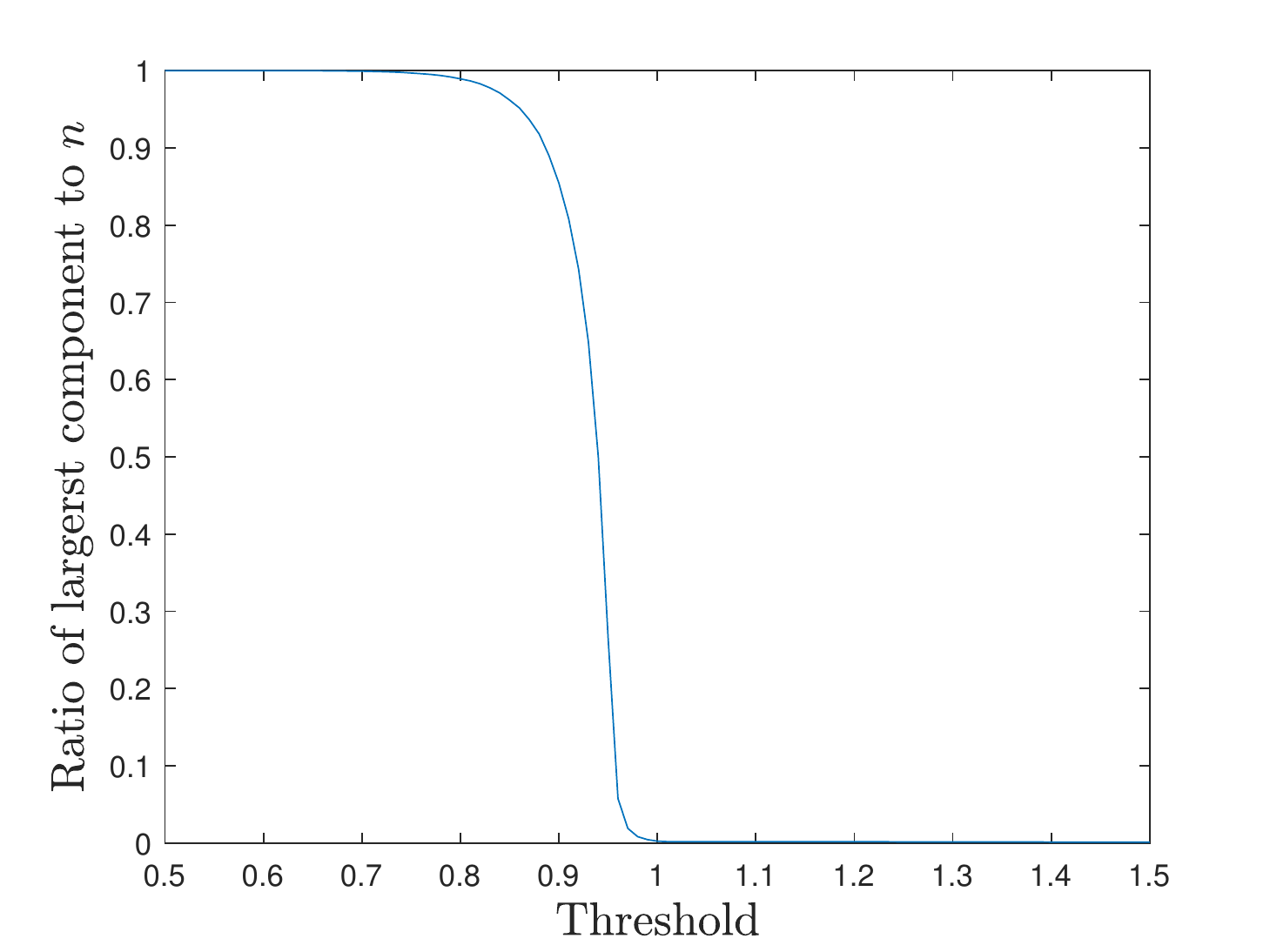}
        \caption{}
    \end{subfigure}
    \begin{subfigure}[t]{0.32\textwidth}
        \centering
        \includegraphics[width=\linewidth]{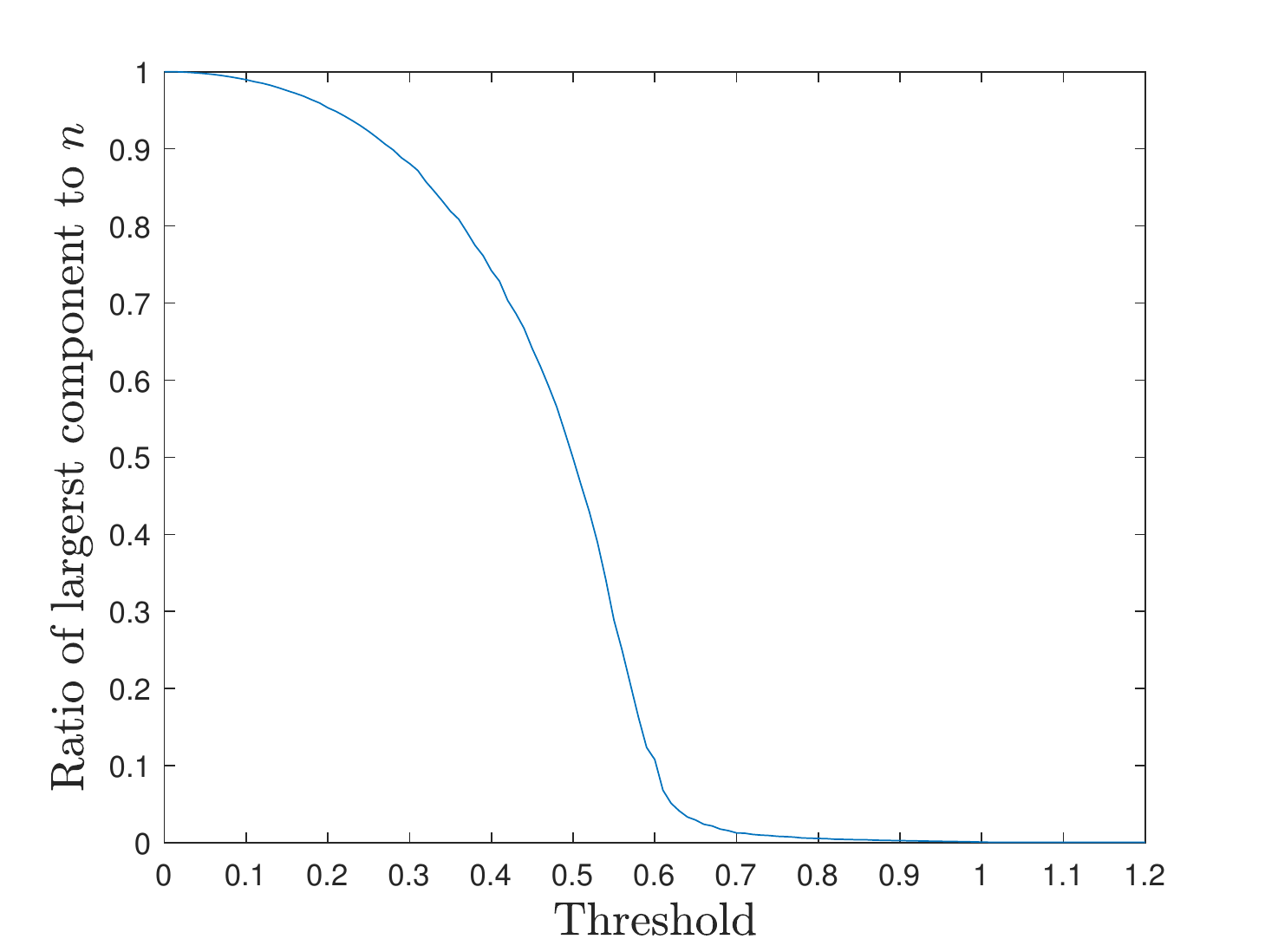}
        \caption{}
    \end{subfigure}
    \begin{subfigure}[t]{0.32\textwidth}
        \centering
        \includegraphics[width=\linewidth]{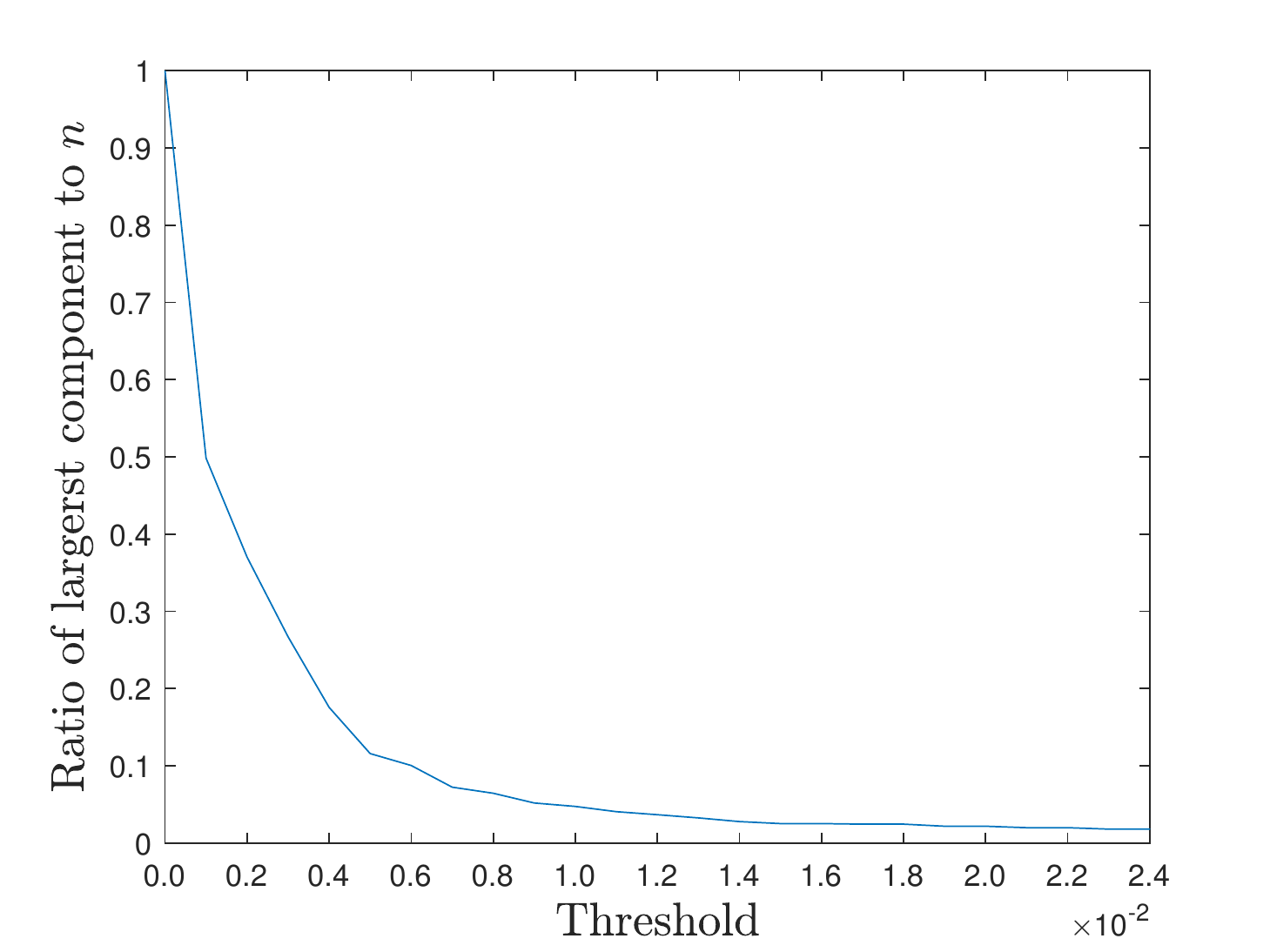}
        \caption{}
    \end{subfigure}
    \caption{Scaling of the largest connected component in a tie-decay network from the model of Ahmad et al. \cite{ahmad2018tie} versus the threshold $g$ for different values of the decay rate $\alpha$. In our simulations, we have $n = 2000$ nodes; decay parameters of (a) $\alpha = 0.001$, (b) $\alpha = 0.01$, and (c) $\alpha = 0.1$; a total simulation time of $T=1000$, and an interaction probability of $p=10^{-5}$ (so $\lambda =Tp=0.01$). Each plot is a mean over 200 instantiations. Using Eq.~\eqref{eq:crit_threshold}, we calculate the critical thresholds $g_{\text{crit}}$ to be (a) $0.9592$, (b) $0.6345$, and (c) $0.0106$.
    }
    \label{fig:crit_threshold}
\end{figure*}
}


\subsection{A simplified version of the back-to-unity model of Jin et al. \cite{jin2001}} 
\label{back to unity growth model}

\begin{figure}[h!]
    \centering
  \includegraphics[width=0.5\textwidth]{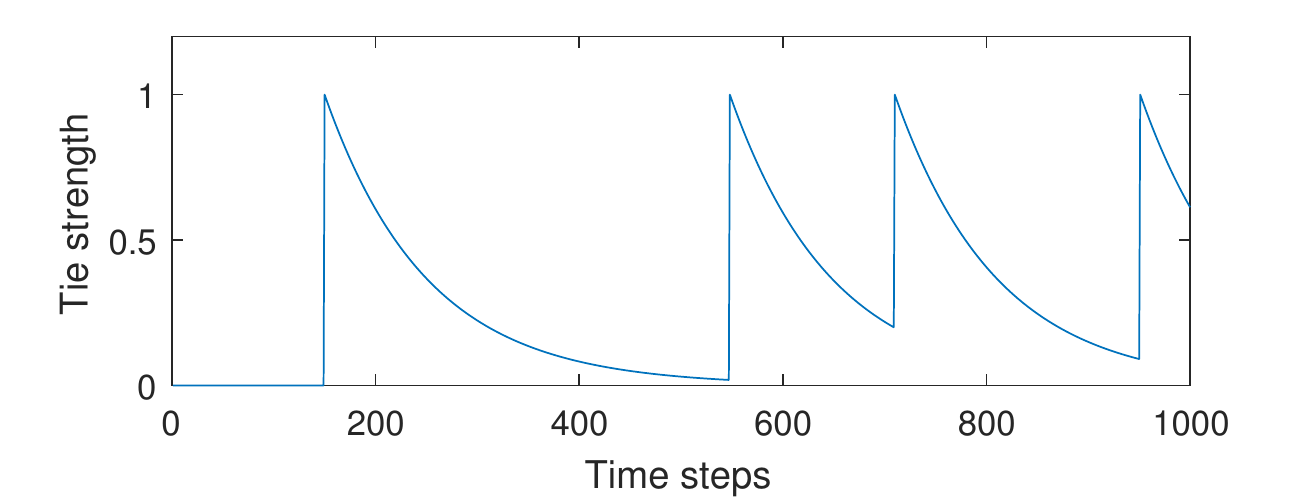}
  \caption{An illustration of tie-decay dynamics of the back-to-unity model of Jin et al. \cite{jin2001}. The tie strength between two nodes resets to $1$ if they interact during a time step. In the depicted simulation, there are $n =1000$ nodes, a decay rate of $\alpha=0.01$, an interaction probability of $p=0.003$, and $T = 1000$ time steps. The vertical axis shows the tie strength of one edge. Four interactions occur between the two nodes that are incident to this edge.
}
  \label{fig:typical_back_to_unity}
\end{figure}

Jin et al. \cite{jin2001} considered a type of tie-decay model (although they did not use that terminology) in which an interaction resets the strength of a tie between two nodes to $1$, instead of increasing the tie strength by $1$ (as in Eq.~\eqref{eq:update_rule}). Consequently, the tie strength of each edge is always bounded above by $1$. In Figure \ref{fig:typical_back_to_unity}, we show an illustrative example of the tie-decay dynamics for the back-to-unity model of \cite{jin2001}.

In their back-to-unity model, Jin et al. \cite{jin2001} set a threshold $g \in (0,1]$ on the tie strength, and they interpreted edges with a tie strength of at least $g$ as active. In their paper, Jin et al. examined the evolution of model friendship networks using numerical simulations. The main assumption in \cite{jin2001} is that a pair of people are more likely to meet with each other when they have common friends than when they do not. Each time two people meet, the tie strength of the edge between them resets to $1$. When they are apart, tie strength between them decreases exponentially. Jin et al. also included an upper bound for the number of active friends that one person can have simultaneously. Using their model, Jin et al. sought to achieve insights into the formation of social networks, and they supposed that a community forms in a network concomitantly with the formation of a GCC. 

In our discussion, we modify (and simplify) the back-to-unity model by dropping (1) the assumption that the chance that two people meet each other depends on their number of their mutual friends and (2) the upper bound on the number of friendships. With this simplified model, we can make some analytical progress. Given an interaction probability $p$ and the threshold $g$, we derive a closed-form expression for the criterion of the emergence of a GCC. 

The long-time behavior of the $n^{\text{th}}$ moment of the tie strength is
\begin{equation}
	\lim_{t\rightarrow \infty} \mathbb{E}[{s_t}^n]  = \frac{p}{1-\sigma^n (1-p)}\,,
\end{equation} 
where we recall that $\sigma = e^{-\alpha}$.

In the time-independent ER random-graph model $G(n,p)$, there is a GCC when
\begin{equation} \label{eq:GCC_criteria}
	p \geq \frac{1+\varepsilon}{n}
\end{equation}
for all $\varepsilon>0$, because there is a phase transition for the emergence of the GCC when $\varepsilon = 0$. When \eqref{eq:GCC_criteria} holds, then with high probability, there is a single GCC and all other components have size $O(\log(n))$ \cite{erdosrenyi1960evolution}. 

\begin{figure*}[t!]
    \centering
    \begin{subfigure}[t]{0.45\textwidth}
        \centering
        \includegraphics[width=\linewidth]{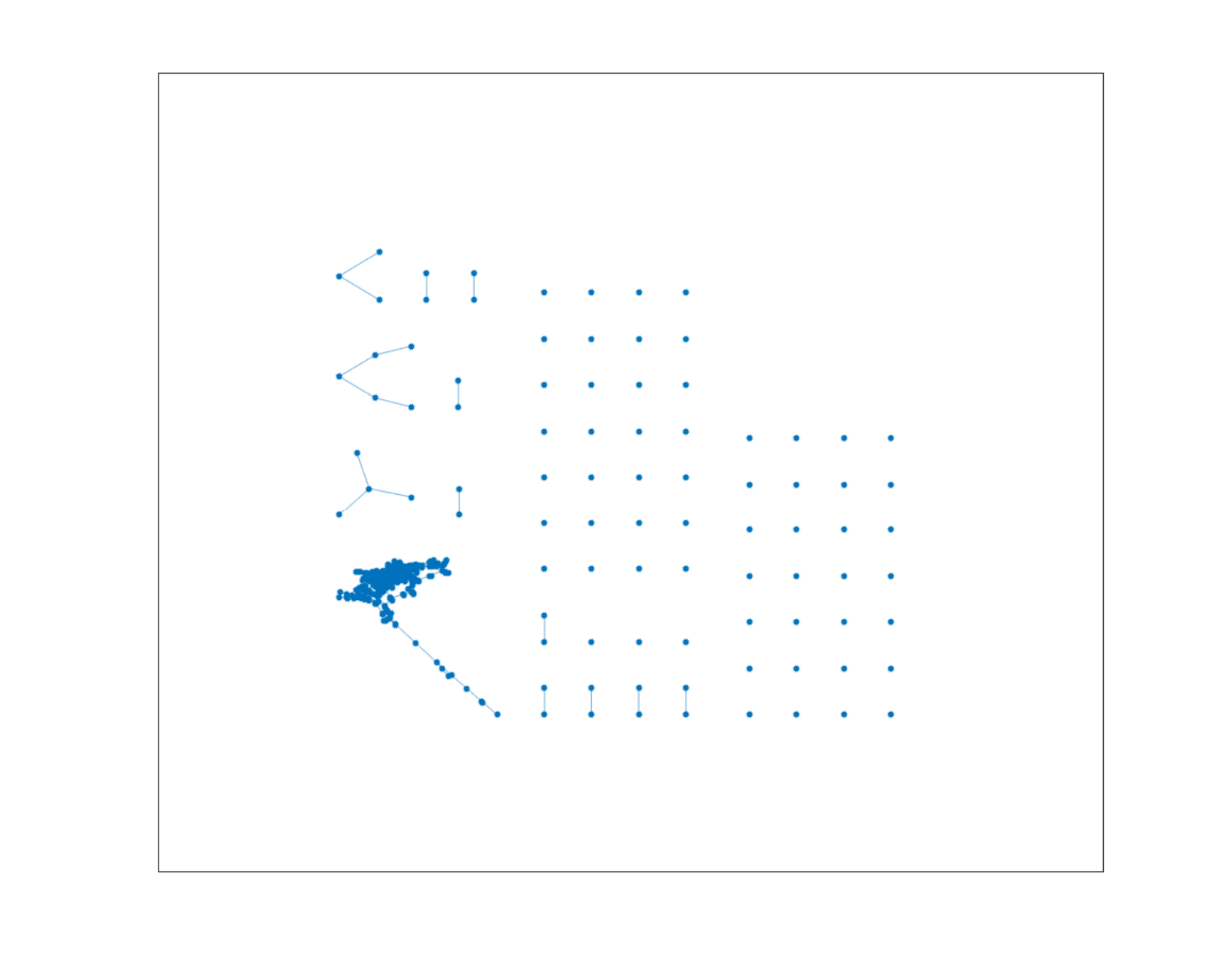}
        \caption{}
    \end{subfigure}
    \begin{subfigure}[t]{0.45\textwidth}
        \centering
        \includegraphics[width=\linewidth]{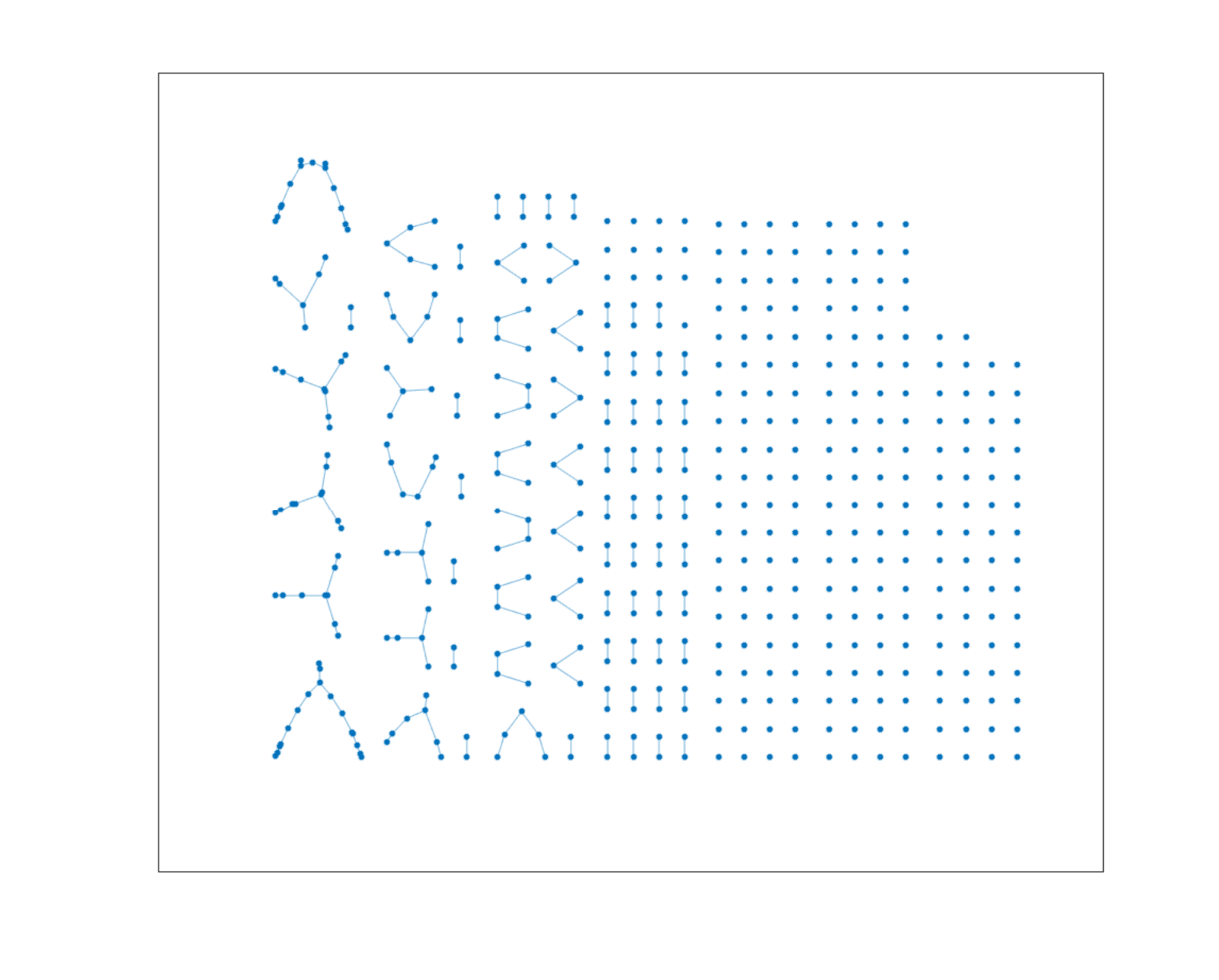}
        \caption{}
    \end{subfigure}
    \caption{Presence versus absence of a GCC in the modified back-to-unity model. In each panel, we show all components of a network from a single simulation. In both simulations, we use $n=1000$ nodes, an interaction probability of $ p= \frac{1}{1.1n}$, a decay parameter of $\alpha =0.01$, and $T = 3000$ time steps. (a) We set the threshold to be $g=0.95$, which yields $\mathbf{P}(s\geq g) \approx 0.0054 > {1}/{n} =0.001$. Therefore, there is a GCC. (b) We set the threshold to be $g=0.995$, which yields $\mathbf{P}(s\geq g) \approx 9.09\times 10^{-4} < {1}/{n}=0.001$. Therefore, there is no GCC.} \label{fig:backtounity_GCC}
\end{figure*}

Because the nodes are indistinguishable from each other, we examine the probability that the strength of a particular edge surpasses the threshold:
\begin{equation} \label{side}
	\mathbf{P}(s \geq g) = 1-\mathbf{P}(s < g)\,.
\end{equation}
We compute the probability on the right-hand side of \eqref{side} as follows. We know that $s$ cannot reset to $1$ in the last step, as otherwise $s=1 \geq g$. Similarly, $s$ cannot reset to $1$ in the last $q$ steps, because otherwise it will not have enough time to decay to some value that is smaller than $g$. Using this argument, we see that the condition that $q$ needs to satisfy is
\begin{equation*}
	e^{-\alpha q} < g\,,
\end{equation*}
which we can express as
\begin{equation*}
	q \geq \Big  \lceil - \frac{\ln(g)}{\alpha}\Big \rceil \,,
\end{equation*}
where $\lceil \theta \rceil$ is the ceiling function of $\theta$ (i.e., the smallest integer that is at least as large as $\theta$).
The probability for this to occur is $(1-p)^q$, so
\begin{align}
	\mathbf{P}(s \geq g) &= 1-\mathbf{P}(s < g) \nonumber \\
	 &= 1-(1-p)^{\Big  \lceil - \frac{\ln(g)}{\alpha}\Big \rceil}\,. \label{eq:bounded_prob}
\end{align}
By the same argument, a GCC exists if 
\begin{equation}
	\mathbf{P}(s \geq g)  = 1-(1-p)^{\Big  \lceil - \frac{\ln(g)}{\alpha}\Big \rceil}  > \frac{1}{n}\,. \label{bounded_prob}
\end{equation}
Because $g\in (0,1]$ and the decay parameter is $\alpha>0$, it follows that $\Big\lceil - \frac{\ln(g)}{\alpha}\Big \rceil >0$. Therefore, we see that if $p>{1}/{n}$, then we are guaranteed that there is a GCC unless $g=1$. That is, $p>{1}/{n}$ is a sufficient condition for the existence of a GCC. Recall that this is also the condition for the existence of a GCC in Eq.~\eqref{eq:GCC_criteria}. Therefore, the criterion for the existence of a GCC in a network that one constructs from the modified back-to-unity model is stricter than that for an ordinary ER $G(n,p)$ graph. In Figure \ref{fig:backtounity_GCC}, we illustrate the presence and absence of a GCC in a network with back-to-unity interactions. Our analytical result in Eq.~\eqref{bounded_prob} agrees with our numerical computations.

\begin{figure*}
\begin{subfigure}[t]{0.32\textwidth}
        \centering
        \includegraphics[width=\linewidth]{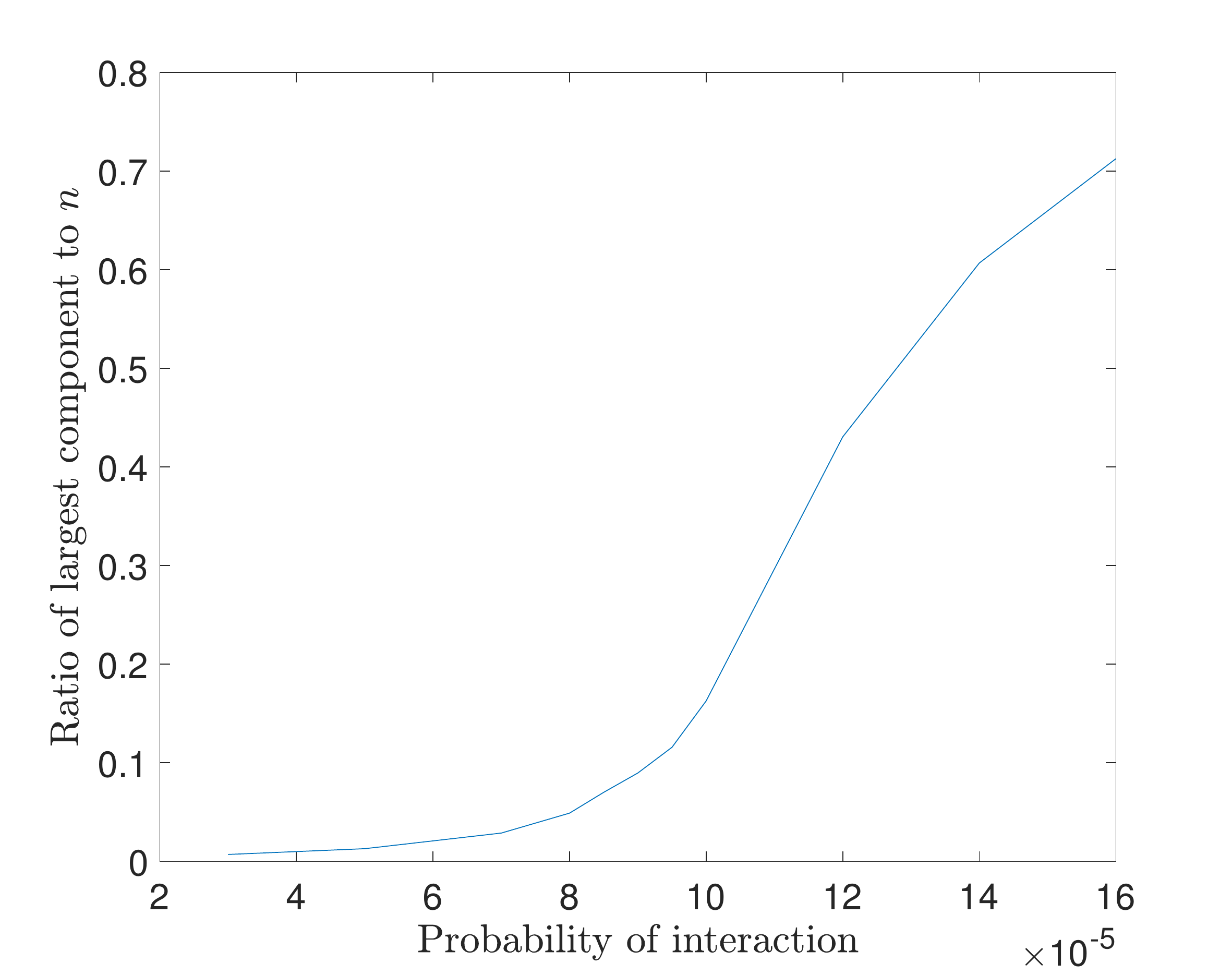}
        \caption{}
    \end{subfigure}
    \begin{subfigure}[t]{0.32\textwidth}
        \centering
        \includegraphics[width=\linewidth]{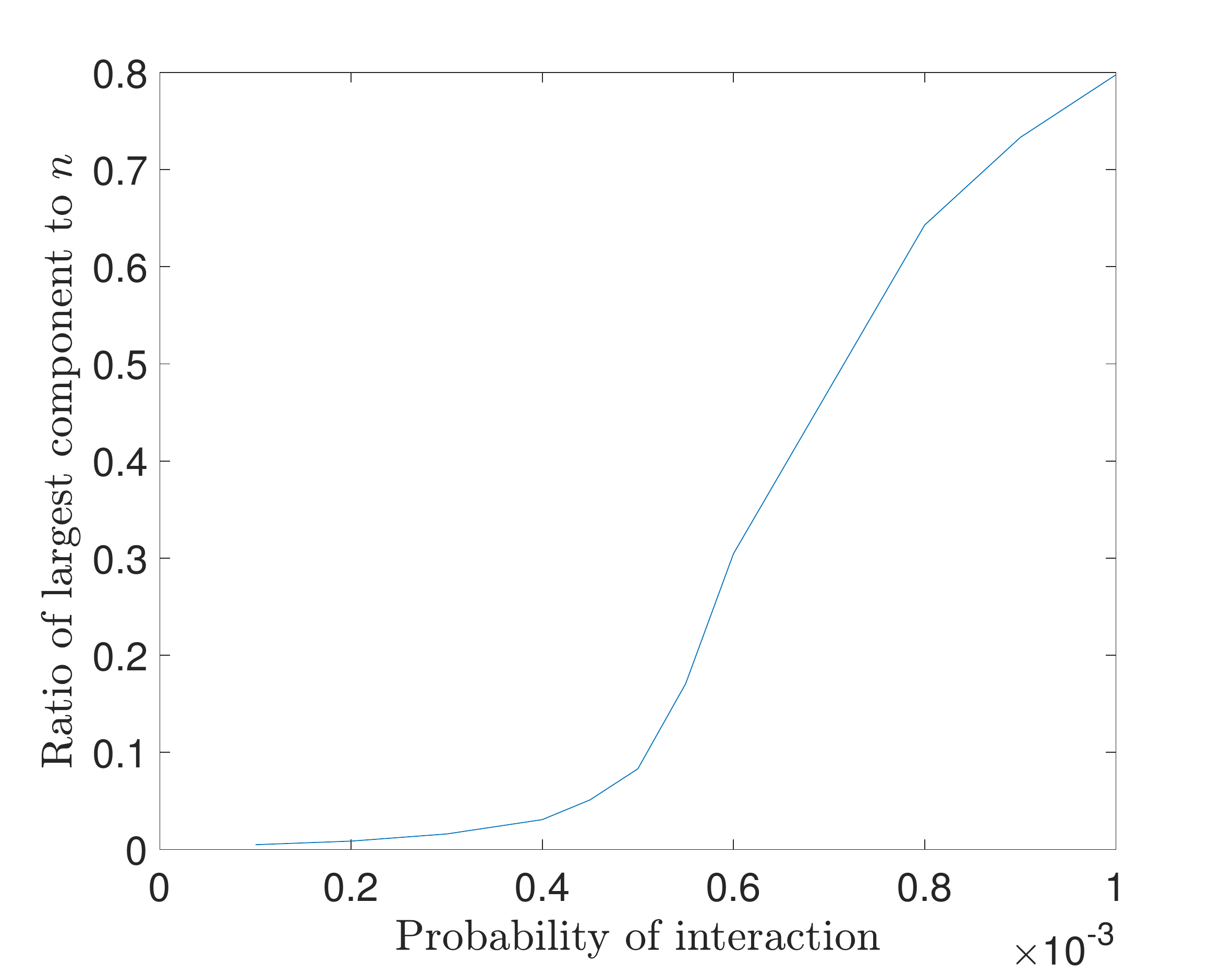}
        \caption{}
    \end{subfigure}
    \begin{subfigure}[t]{0.32\textwidth}
        \centering
        \includegraphics[width=\linewidth]{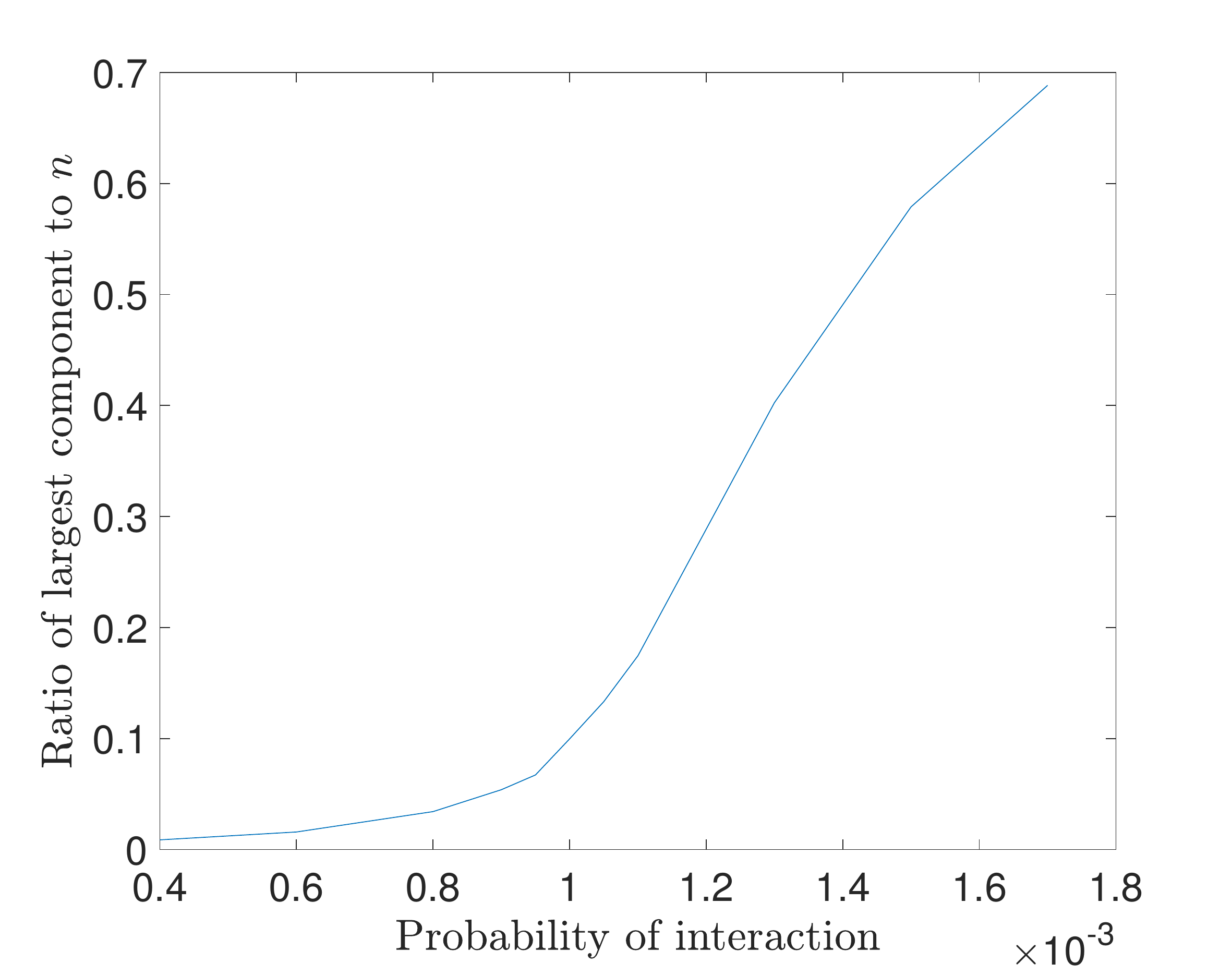}
        \caption{}
    \end{subfigure}
    \caption{Scaling of the GCC in a network that we construct using the back-to-unity model with interaction probability $p$. In our simulations, there are $n=1000$ nodes; a decay parameter of (a) $\alpha=0.01$, (b) $\alpha=0.1$, and (c) $\alpha = 1$; and a threshold of $g=0.9$. For each value of $p$, we take a mean of our results over $250$ realizations. Each realization has a run time of $T=500$. Using Eq.~\eqref{bounded_prob}, we calculate the critical probabilities $p_{\text{crit}}$ to be (a) $9\times 10^{-5}$, (b) $0.5\times 10^{-3}$, and (c) $1\times 10^{-3}$. }\label{fig:ratio_p}
\end{figure*}

We also investigate numerically how the size of the GCC (if there is one) in a network that we construct from the modified back-to-unity model varies with the interaction probability $p$ when we fix all other parameters. We show the results of our numerical computations in Figure \ref{fig:ratio_p}. Based on Eq.~\eqref{bounded_prob} and our parameter values, we calculate that the critical probabilities $p_{\text{crit}}$ for a GCC to emerge are $p_{\text{crit}} \approx 9\times 10^{-5}$, $p_{\text{crit}} \approx 0.5\times 10^{-3}$, and $p_{\text{crit}} \approx 1\times 10^{-3}$ for decay rates of $\alpha = 0.01$, $\alpha = 0.1$, and $\alpha = 1$, respectively. As we can see with our simulations, we expect to observe a phase transition at $p=p_{\text{crit}}$.


\subsection{Diffusion model of tie strengths}
\label{diffusion model}

Another continuous-time model, which we introduce in the present paper, is a toy model of a tie-decay network based on diffusion. At each time step, each entity is equally likely either to interact with some entity or to not do anything. Each interaction that occurs between a pair of nodes is independent of all other pairs (i.e., all other edges), so the strength of each tie changes independently of all other ties. This implies that, at each time step, there is an equal probability ${1}/{2}$ for the tie strength of each edge to increase or decrease by the factor $\exp(\pm \delta x)$, where $\delta x$ is small. We assume that the tie strength of each edge starts at $\exp(0)=1$. We then show that, as time progresses, we can model the tie strength by a linear diffusion equation, similar to how one derives a diffusion equation from a symmetric random walk. In Section \ref{bounded convection diffusion model}, we will generalize our diffusion model to include both diffusion and convection.

\begin{figure*}[htbp]
    \centering
    \begin{subfigure}[htbp]{0.45\textwidth}
        \centering
        \includegraphics[width=\linewidth]{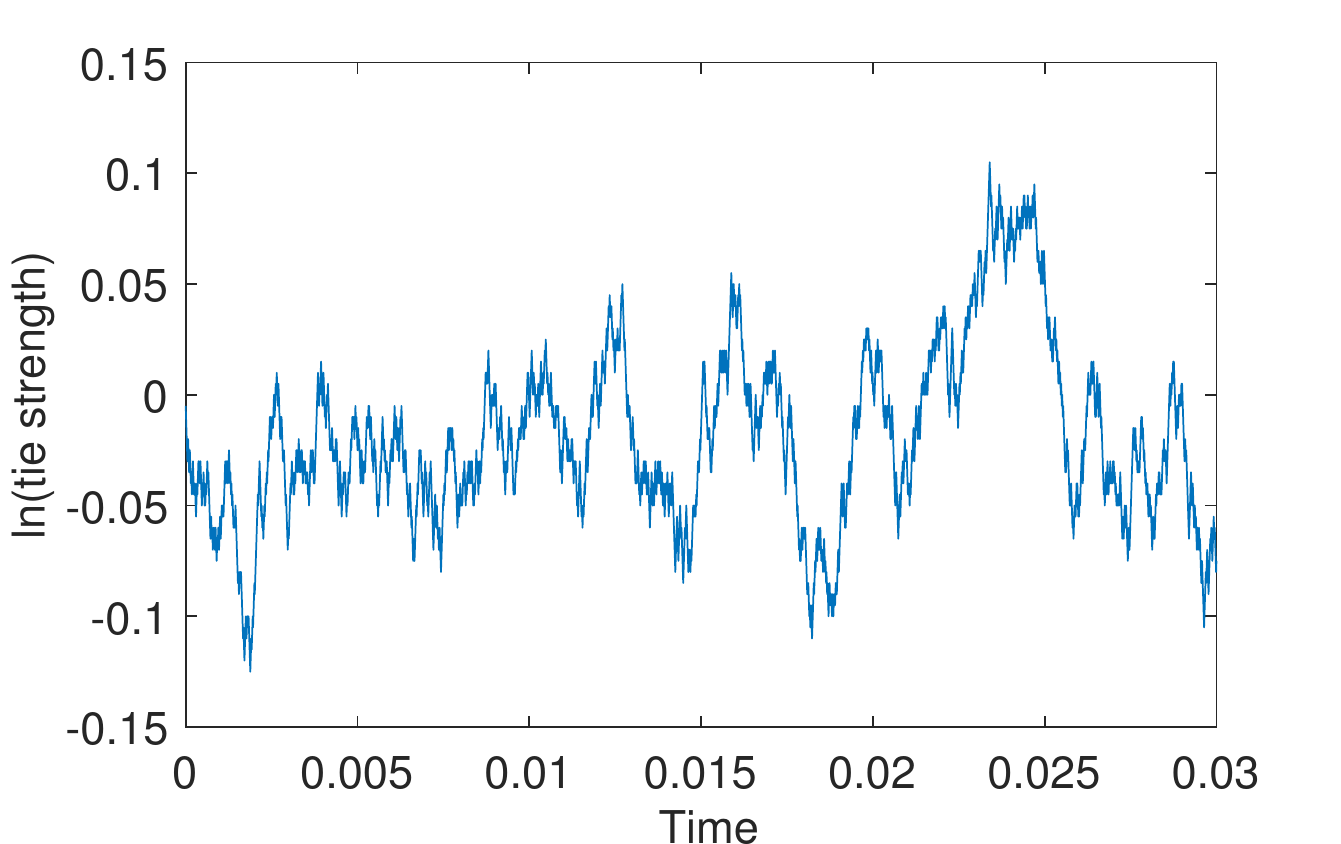}
        \caption{}
    \end{subfigure}%
    ~ 
    \begin{subfigure}[htbp]{0.45\textwidth}
        \centering
        \includegraphics[width=\linewidth]{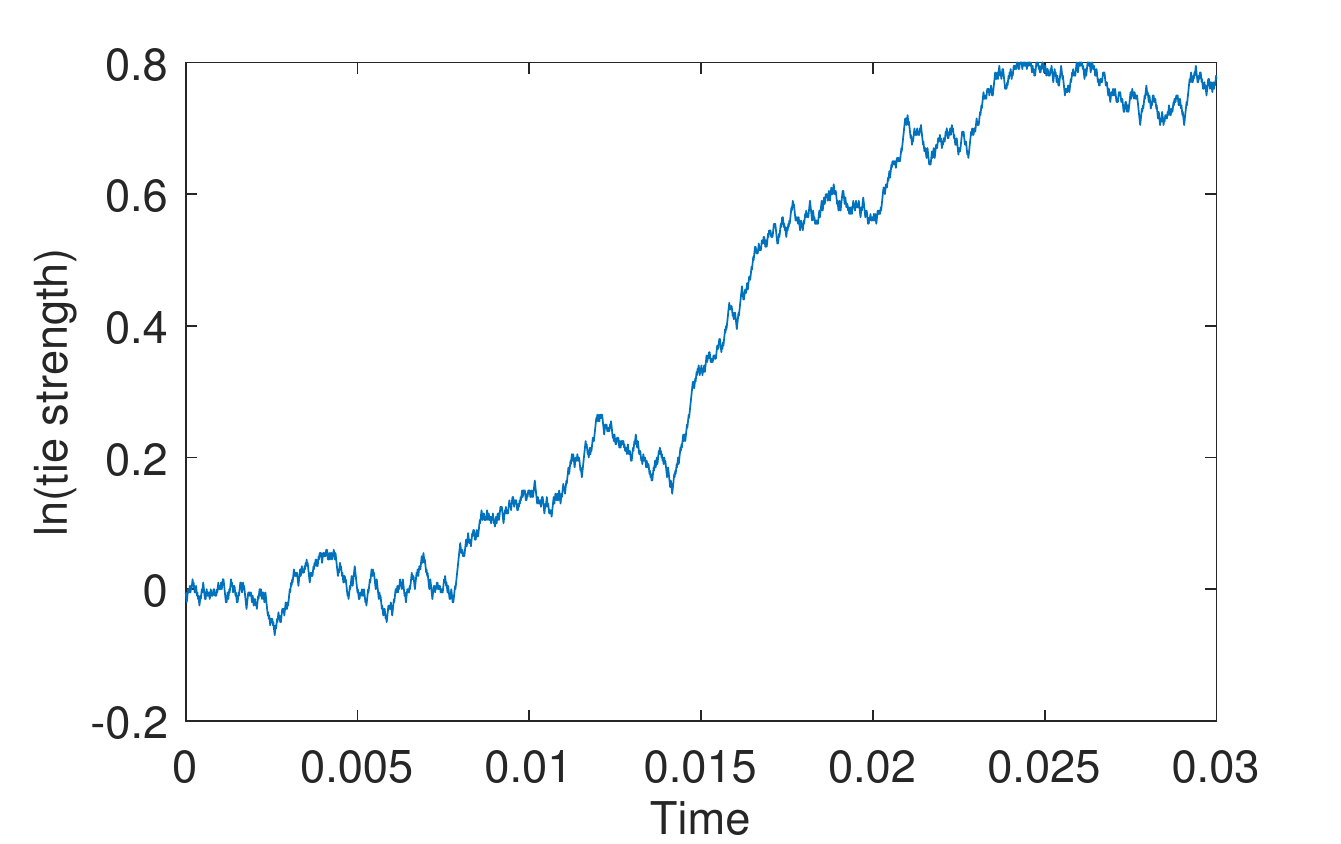}
        \caption{}
    \end{subfigure}
    \caption{An illustration of tie-strength dynamics for (a) our diffusion model and (b) our bounded convection--diffusion model. In both panels, we use a spatial step of $\delta x = 5\times 10^{-3}$, a time step of $\delta t = 10^{-5}$, and a simulation time of $T = 0.03$. For panel (b), the convection parameter is $\beta = 5$ and the upper bound of the tie strength is $w = 0.8$. The vertical axis in each panel shows the natural logarithm of the tie strength of a single edge. In contrast to our simulations of the Ahmed et al. tie-decay model and the modified back-to-unity model, the time step $\delta t \neq 1$. The diffusion model has a first-order error in time (as well as in space), so we need the time step to be small. 
   }\label{fig:diffusion_growth}
\end{figure*}

Because the tie strength of each edge changes independently of other edges, we examine the dynamics of a single edge. Let $u(x,t)$ denote the probability that the tie strength of a chosen edge at time $t$ is given by $\exp(x)$. We write the master equation
\begin{equation}
	    u(x,t) = \frac{1}{2}[u(x-\delta x,t-\delta t) + u(x + \delta x,t-\delta t)]\,,
\end{equation}
which we rearrange to obtain
\begin{align}\label{rearrange}
    &u(x,t+\delta t)-u(x,t) \\
    &=\frac{1}{2}[u(x+\delta x,t)-u(x,t)- (u(x,t)-u(x-\delta x,t))]\,. \nonumber
\end{align}
We now approximate finite differences as derivatives in Eq.~\eqref{rearrange}. Specifically, we take $\delta t \rightarrow 0$ and $\delta x \rightarrow 0$, while supposing that $\frac{(\delta x)^2}{\delta t}$ does not go to $0$, to yield the PDE
\begin{align} \label{eq:diffusion}
    \frac{\partial u}{\partial t} \delta t &= \frac{1}{2}\left((\delta x)^2 \frac{\partial^2u}{\partial x^2}\right) \nonumber\\
   \implies u_{t} &= \frac{1}{2}\frac{(\delta x)^2}{\delta t} u_{xx} + O(\delta x)+O(\delta t)\,,
\end{align}
where we use the notation $u_{t} \equiv \frac{\partial u}{\partial t}$ (and analogously for spatial derivatives). The initial condition in \eqref{eq:diffusion} is
\begin{equation} \label{eq:delta_initial}
    u(x,0) = \delta (x)\,,
\end{equation}
where $\delta(x)$ is the Kronecker delta function (and should not be confused with $\delta x$, an infinitesimal change in the variable $x$). This initial condition implies that, at time $0$, the tie strength of the chosen edge is $\exp{(0)}$ with probability $1$. Equation \eqref{eq:diffusion} and Eq.~\eqref{eq:delta_initial} constitute a diffusion equation with a delta-mass initial condition. With this initial condition, we can solve this equation both numerically and analytically. If we define $D = \frac{1}{2}\frac{(\delta x)^2}{\delta t}$, we obtain the similarity solution \cite{evans1997partial}
\begin{equation}
    	u(x,t) = \frac{1}{\sqrt{4 \pi Dt}}\exp\left(-\frac{x^2}{4Dt}\right)\,.\label{eq:soln_Gaussian}
\end{equation}
Therefore, the tie strength in the diffusion model spreads out over time in a Gaussian manner. 


\subsection{Bounded convection--diffusion model of tie strengths}
\label{bounded convection diffusion model}

We now modify the diffusion model in Section \ref{diffusion model} by supposing that there is a preference for tie strengths to grow over time. At each time step, there is a probability of $({1}/{2}+\Delta)$ for a tie strength to grow by a factor of $\exp(\delta x)$ and a probability of $({1}/{2}-\Delta )$ for it to decay by a factor of $\exp(-\delta x)$. We also suppose that $\Delta$ is small. We view the growth pattern of the tie strength as a one-dimensional (1D) random walker that has a preference to move in the positive direction. The associated master equation is
\begin{equation} \label{eq:convec_diff_model}
    u(x,t+\delta t) = \left(\frac{1}{2}+\Delta \right)u(x-\delta x,t) + \left(\frac{1}{2}-\Delta \right) u(x+\delta x,t)\,. 
\end{equation}
Following a similar procedure as with the diffusion model in Section \ref{diffusion model}, we derive the equation
\begin{equation} \label{eq:convec_diff}
    u_{t} = ku_{xx} - 4\beta ku_{x}+O(\delta x) + O(\delta t)\,, 
\end{equation}
where we assume that $\frac{(\delta x)^2}{2 \delta t}\rightarrow \text{constant} = k$ and $\frac{\Delta}{\delta x} \rightarrow \text{constant} = \beta$. In this regime, we obtain a convection--diffusion equation, with the delta-mass initial condition \eqref{eq:delta_initial}. 

To prevent our random walker from escaping to infinity, we enforce that a tie strength has an upper bound $W$. Specifically,
\begin{equation} \label{eq: convec_diff_bound}
    u(x,t) = 0 \quad \text{for all} \quad x> w\,, 
\end{equation}
where $w = \ln{W}$. Equation \eqref{eq: convec_diff_bound} is a linear diffusion equation in a moving frame. We make the change of variables $(x,t)\rightarrow (\xi,t)$, where $\xi = x-4\beta k t$. By the chain rule, Eq.~\eqref{eq:convec_diff} becomes 
\begin{equation}  \label{eq:diff_moving_frame}
    	u_{t} = ku_{\xi \xi}\,,
\end{equation}
which is the usual diffusion equation. 

Together with conservation of probability, Eq.~\eqref{eq: convec_diff_bound} enforces a boundary condition in our scheme for our numerical computations of \eqref{eq:convec_diff}. Using a forward-time, central-difference scheme gives
\begin{align} \label{eq:scheme}
    \frac{u^{i+1}_j-u^i_j}{\delta t} &= k\frac{u^i_{j+1}-2u^i_j+u^i_{j-1}}{(\delta x)^2} - 4\beta k \frac{u^i_{j+1}-u^i_{j-1}}{2 \, \delta x}\,, \nonumber \\
    u^{i+1}_j &= a u^i_{j-1}+b u^i_j + c u^i_{j+1}\,,
\end{align}
where the superscript $i$ on $u$ indicates the time discretization, the subscript $j$ on $u$ indicates the spatial discretization, and 
\begin{align} \label{eq:abc} 
    	a &= k\frac{\delta t}{(\delta x)^2} + 2\beta k \frac{\delta t}{\delta x}\,, \nonumber\\
    	b &= 1-2k\frac{\delta t}{(\delta x)^2}\,, \\
    	c &= k \frac{\delta t}{(\delta x)^2} - 2k\beta \frac{\delta t}{\delta x}\,. \nonumber
\end{align}
Inserting the expressions for $k$ and $\beta$ into \eqref{eq:abc} yields $a = \frac{1}{2}+\Delta$, $b=0$, and $c=\frac{1}{2}-\Delta$. Inserting these values into our numerical scheme in \eqref{eq:scheme} yields
\begin{equation}
     u^{i+1}_j = \left(\frac{1}{2}+ \Delta \right) u^i_{j-1} + \left(\frac{1}{2}- \Delta \right) u^i_{j+1}\,,
\end{equation}
which is equivalent to Eq.~\eqref{eq:convec_diff_model}. This indicates that the numerical scheme in Eq.~\eqref{eq:scheme} successfully describes the evolution of the tie strength of an edge if one is sufficiently far away from the boundary. By the minimum principle and the infinite speed of wave propagation in our linear convection--diffusion equation \cite{evans1997partial}, Eq.~\eqref{eq:convec_diff} and Eq.~\eqref{eq:diff_moving_frame} give a nonzero solution at the boundary for any $t>0$, but our discrete system in Eq.~\eqref{eq:convec_diff_model} has a nonzero solution at the boundary only after some finite time. 

In principle, $u$ can take a negative value of arbitrarily large magnitude. However, to implement the numerical scheme \eqref{eq:scheme}, we have to use a finite interval. As we discussed above, we impose an upper bound on $w$. Although the solution to the convection--diffusion equation \eqref{eq:convec_diff} has an infinite propagation speed, our discrete model has a finite propagation speed $v = \frac{\delta x}{\delta t}$. Therefore, we can choose a lower bound $-L$ (with $L \in \mathbb{R}_{> 0}$), such that $L \geq \frac{T}{v}$. That is, at $t=T$, we have $u(x,T)=0$ for all $x\leq -L$. Therefore, $u_j^i = u(x_j,t_i)$, with our spatial discretization given by $\{x_0=-L,x_1,\ldots,x_N=w\}$ and our time discretization given by $\{t_0=0,t_1,\ldots,t_{N_T}=T\}$.

We derive boundary conditions by requiring conservation of mass:
\begin{equation} \label{eq:normalization_condition}
	    \sum^N_{j=1}u^{i+1}_j  = \sum^N_{j=1}u^{i}_j\,. 
\end{equation}
Combining Eqs.~\eqref{eq:scheme} and \eqref{eq:normalization_condition} yields
\begin{align} \label{eq:boundary_condition}
	    u^{i+1}_1 &= u^i_1 (1-a) + u^i_2(1-a-b)\,, \nonumber\\
	    u^{i+1}_N &= u^i_N(1-c) + u^i_{N-1}(1-b-c)\,. 
\end{align}

We now examine the boundary at $x=w$. From Eq.~\eqref{eq:boundary_condition}, the boundary condition on the right (which we derive from conservation of mass) is 
\begin{equation} \label{eq:numerics_boundary}
    u^{i+1}_N = \left(\frac{1}{2}+\Delta \right)u^i_N + \left(\frac{1}{2}+\Delta \right)u^i_{N-1}\,.  
\end{equation}
However, our model requires that the tie strength of an edge cannot exceed some threshold $w$. Therefore, whenever the tie strength of an edge reaches $w$, we require at the next time step that it either remains at $w$ or decays to $w-\delta x$. Similarly, if the tie strength of an edge is $w$ at time $t$, then the tie strength of that edge at time $t-\delta t$ is either $w-\delta x$ or $w$. If, at some time, the tie strength $x$ is smaller than $w$ but becomes $x+\delta x \geq w$ at the next time step, we always set the new tie strength to $w$. Mathematically, the above boundary conditions translate to 
\begin{align} \label{eq:bc_randomwak}
    u(w,t+\delta t) &=\left(\frac{1}{2}+ \Delta \right) u(w,t) + \left(\frac{1}{2}+ \Delta \right)u(w-\Delta x,t)\,, \nonumber \\
     u^{i+1}_N &= \left(\frac{1}{2}+\Delta \right)u^i_N + \left(\frac{1}{2}+\Delta \right)u^i_{N-1}\,. 
\end{align}
Consequently, the natural boundary condition from the model is equivalent to the boundary condition that we impose on our numerical scheme \eqref{eq:scheme} based on conservation of mass. 

After choosing the upper and lower bounds for our numerical computations, we implement our numerical scheme \eqref{eq:scheme} by building a transition matrix from Eq.~\eqref{eq:scheme} and Eq.~\eqref{eq:boundary_condition}. This matrix is a tridiagonal matrix in which every column sums to $1$, so it is a stochastic matrix and there always exists an eigenvector with eigenvalue $1$. This transition matrix is a positive stochastic matrix if $\Delta < \frac{1}{2}$. By the Perron--Frobenius theorem, the eigenspace of the unit eigenvalue is spanned by one vector, which is the stationary state. 

At steady state, $u(x,t) = u(x,t+\delta t)$, so it follows that $u^{i+1}_j = u^i_j$. Inserting this relation into Eq.~\eqref{eq:numerics_boundary} yields
\begin{align} \label{eq:numerical_bc}
    u_N &= \left(\frac{1}{2} + \Delta \right)u_{N-1} +  \left(\frac{1}{2}+ \Delta \right)u_{N}\,,
\end{align}
which implies that
\begin{align}    
    \frac{u_N}{u_{N-1}} &= \frac{\frac{1}{2}+\Delta }{\frac{1}{2}-\Delta}\,. 
\end{align} 
Similarly, away from the boundary, 
\begin{align}
    u_{N-1} &= \left(\frac{1}{2}+ \Delta \right)u_{N-2} +  \left(\frac{1}{2}- \Delta \right)u_{N} \nonumber \\
    &= \left(\frac{1}{2}+ \Delta \right)u_{N-2} +  \left(\frac{1}{2}- \Delta \right)\frac{\frac{1}{2}+\Delta }{\frac{1}{2}-\Delta}u_{N-1} \nonumber\\
    &= \left(\frac{1}{2}+ \Delta \right)u_{N-2} +  \left(\frac{1}{2}+ \Delta \right)u_{N-1}\,,
\end{align}
which implies that
\begin{equation}
    \frac{u_{N-1}}{u_{N-2}} = \frac{\frac{1}{2}+\Delta}{\frac{1}{2}-\Delta}\,. 
\end{equation}
By induction, we obtain
\begin{equation} \label{eq:ratio_adjacent}
     \frac{u_{j}}{u_{j-1}} = \frac{\frac{1}{2}+\Delta}{\frac{1}{2}-\Delta}\,. 
\end{equation}
From conservation of mass, 
\begin{equation} \label{eq:conservation_of_mass} 
    \int_\mathbb{R} u(x,t) dx = \text{constant}\,, 
\end{equation}
which yields 
\begin{align} \label{eq:continuous_bc}
    u_{x}(w) &= 4\beta u(w)\,,
\end{align}
where we take $u\rightarrow 0$ and $u_{x}\rightarrow 0$ as $x\rightarrow -\infty$ based on our numerical computations. From the continuous model \eqref{eq:continuous_bc}, we thus obtain the following boundary conditions for our numerical computations: 
\begin{align} \label{eq:continuous_bc_scheme}
    \frac{u_N - u_{N-1}}{\delta x} &= 4\beta u_N\,, \nonumber\\
    u_N &= \frac{u_{N-1}}{1-4\Delta}\,.
\end{align}
The boundary conditions in \eqref{eq:continuous_bc_scheme} are not exactly the same as those that we derived directly from the numerical conservation of mass in Eq.~\eqref{eq:normalization_condition} or from the network model in Eq.~\eqref{eq:bc_randomwak}. However, Eq.~\eqref{eq:numerical_bc} and Eq.~\eqref{eq:continuous_bc_scheme} agree to first order in $\Delta$.

\begin{figure}[h!]
  \includegraphics[width=0.5\textwidth,height=0.4\textwidth]{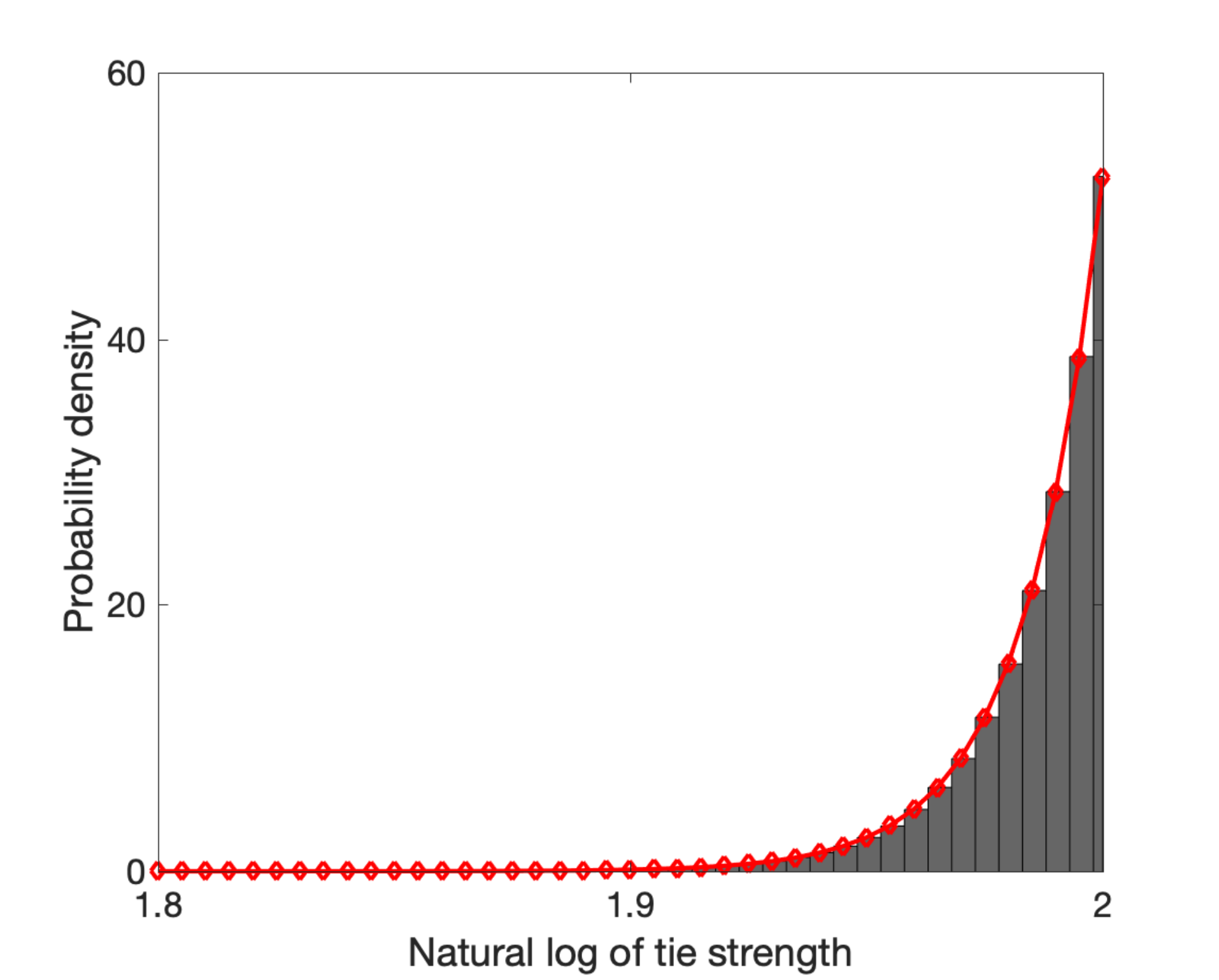}
  \caption{Comparison of our numerical scheme (red curve) from Eq.~\eqref{eq:scheme} and a mean of Monte Carlo simulations over 100 realizations of our network model (gray blocks) with a bounded convection--diffusion model of tie strengths. This figure gives the probability distribution of the tie strength of an edge in a network. Because we assume that each edge is independent of all other edges, this distribution applies to the tie strength of each edge in the network. 
   }
  \label{fig:tie_strength_dist}
\end{figure}

Let $\eta = \frac{\frac{1}{2}-\Delta}{\frac{1}{2}+\Delta}$. From the delta-mass initial condition, we have the geometric sum 
\begin{align}
    \sum_{j=1}^N u_i &= \frac{1}{\delta x} \nonumber \\
    &= u_N(1+\eta+\eta^2+ \ldots +\eta^{N-1}) \nonumber \\
    &= u_N \frac{1-\eta^N}{1-\eta} \nonumber \\
    &= u_N\frac{(1-\eta^N)(\frac{1}{2}+\Delta)}{2\Delta}\,. 
\end{align}
Recall that $N = \frac{w+L}{\delta x}$ and $\eta<1$. Therefore, for our asymptotic solution to Eq.~\eqref{eq:scheme} at steady state with boundary conditions \eqref{eq:boundary_condition}, we may take $\eta^N \rightarrow 0$. In this asymptotic limit, we solve for $u_N$ in terms of $\Delta$ and $\beta$ to obtain
\begin{equation} \label{eq:exact_u_N}
    u_N = \frac{2\Delta}{(\frac{1}{2}+\Delta)\delta x} = \frac{2\beta}{(\frac{1}{2}+\Delta)}\,. 
\end{equation}
We implement our numerical scheme \eqref{eq:scheme} and boundary conditions \eqref{eq:boundary_condition} with the parameter values $\delta x=5 \times10^{-3}$, $\delta t=10^{-5}$, $T=0.05$, $\beta = 15$, $w=2$, and $\Delta = 7.5 \times 10^{-2}$. We also run 100 Monte Carlo simulations on a network with $n=2000$ nodes with the mechanism {that we described in the first paragraph of Section \ref{bounded convection diffusion model} and these same parameter values. We average the simulations and show our results in Fig.~\ref{fig:tie_strength_dist}. }

Equation \eqref{eq:exact_u_N} implies that the solution to the PDE \eqref{eq:convec_diff} at the boundary $w$ at stationarity does not depend on the value of $w$. This is pleasing, because there is no particular reason to choose one value of $w$ over another. 

The numerical scheme in \eqref{eq:scheme}, together with boundary conditions \eqref{eq:boundary_condition}, is accurate to first order both in time and in space. Because $u_N$ converges to the steady-state solution $u(w)$ as we decrease $\delta x$, we can make concrete statements about the exact stationary-state solution to the convection--diffusion equation with mass-conserving boundary conditions. The boundary value of $u$ at stationarity is
\begin{equation} \label{eq:exact_soln_boundary}
	 u(w)=4\beta\,. 
\end{equation}
From Eq.~\eqref{eq:ratio_adjacent}, we obtain the following expression for $u(x,t)$ for $x<w$ in the long-time limit (i.e., at stationarity): 
\begin{equation}  \label{eq:analytic_solution}
    u(x,t) = \lim_{m\rightarrow \infty}\left(\frac{\frac{1}{2}-\beta \frac{w-x}{m}}{\frac{1}{2}+\beta \frac{w-x}{m}}\right)^m 4\beta\,.
\end{equation}
One can think of the expression \eqref{eq:analytic_solution} as taking the limit of our numerical scheme \eqref{eq:scheme} as the step size $\delta x$ of our spatial discretization goes to $0$. Additionally, if $u(x,t)$ is a solution to the convection--diffusion equation at stationarity, it follows that
\begin{align} \label{eq:analytic_ratio}
    \frac{u(w,t)}{u(w-\delta x,t)} &= \frac{u(w-\delta x,t)}{u(w-2\,\delta x,t)}\nonumber \\
    &= \frac{u(w-\ell \,\delta x,t)}{u(w-(\ell+1)\,\delta x,t)} 
\end{align}
for $\ell\in \mathbb{Z}_{\geq 0}$. 

From Eq.~\eqref{eq:analytic_ratio}, we write down the solution to Eq.~\eqref{eq:convec_diff} at stationarity. The solution is of the form 
\begin{equation}
    u(x) = C e^{B (x-w)}\,.
\end{equation}
We determine the constants $B$ and $C$ from Eq.~\eqref{eq:conservation_of_mass} and Eq.~\eqref{eq:exact_soln_boundary} to obtain
\begin{equation} \label{eq:exact_stationary_solution}
    u(x) = 4\beta e^{4\beta (x-w)}\,.
\end{equation}
{In the discrete case, the solution is 
\begin{equation}
    u_{N-j} = \frac{4\beta}{1+2\Delta} \exp\left\{\frac{4 j \Delta}{1+2\Delta}\right\}\,, \quad j \in \{0, \ldots, N\}\,.
\end{equation}
Therefore, a necessary (but not sufficient) condition for our convection--diffusion network to not have a GCC is
\begin{equation}
    \Delta < \frac{1}{4n-2}\,.
\end{equation}}
The solution~\eqref{eq:exact_stationary_solution} satisfies the original convection--diffusion equation \eqref{eq:convec_diff}, subject to the conditions in Eqs.~\eqref{eq:conservation_of_mass}, \eqref{eq:exact_soln_boundary}, and \eqref{eq:analytic_ratio}. The formula~\eqref{eq:exact_stationary_solution} provides a way to examine the emergence of a GCC in our convection--diffusion network model in the long-time limit. At stationarity, the probability distribution for the tie strength of edges is biased towards the boundary, so we look near the boundary $x = w$ for potentially interesting behavior. We define a threshold $W_0<W$ and let $w_0 = \ln(W_0)$. Recall that when looking for a GCC in a network, we interpret all edges with tie strengths that are smaller than or equal to the threshold value as inactive. We interpret edges with tie strengths that are larger than the threshold as active. In the continuum limit, assuming that the system \eqref{eq:convec_diff_model} has reached stationarity, we calculate the probability that a particular edge has a tie strength that is larger than the threshold. This probability is 
\begin{equation}
    \mathbf{P} = 1-e^{4\beta(w_0-w)}\,. \label{eq:convec_diff_dist_eqb}
\end{equation}
When $\mathbf{P}>{1}/{n}$, our network has a GCC. A phase transition occurs when $\mathbf{P}={1}/{n}$.

One can also study the stationary states of the convection--diffusion equation \eqref{eq:convec_diff} directly by setting $u_{t}=0$ to obtain an ordinary differential equation (ODE). This gives the same result as Eq.~\eqref{eq:exact_stationary_solution}, provided we use the same initial and boundary conditions.

Another way to look at growth and decay of tie strength in our convection--diffusion network model is by identifying a process as a 1D birth--death chain \cite{levin2017markov}. However, if we use such an approach, we need a state space $\Omega$ that is not bounded from below. A way to do this is to first examine a finite state space and then take the limit as the lower bound goes to negative infinity. A 1D birth--death chain gives another way to derive Eq.~\eqref{eq:ratio_adjacent}. 

An advantage of analyzing our tie-decay temporal network model using ideas from convection--diffusion equations is that it allows us to write down a characteristic time scale for a network to reach a stationary state. From Eq.~\eqref{eq:diff_moving_frame}, we can view the convection--diffusion equation as a diffusion equation in a moving frame. Let $\tau_1$ to be the time for the initial configuration to move to the boundary at $x = w$. This time is given by
\begin{equation*}
	\tau_1 = \frac{w}{4\beta k}\,.
\end{equation*}
Because the solution \eqref{eq:soln_Gaussian} to the diffusion equation is a Gaussian distribution that expands over time, we define $\tau_2$ to be the time scale for the initial configuration to expand until it has a standard deviation of $w$. This time is given by
\begin{equation*}
	\tau_2 =\frac{w^2}{2k}\,.
\end{equation*}
Therefore, the characteristic time scale to reach stationarity when starting from a delta mass initial condition is $\tau = \text{max}\{\tau_1,\tau_2\}$. An interesting observation is that this is similar to determining a time scale based on the P\'eclet number \cite{batchelor1967introduction}, which measures the relative strengths of convection and diffusion. If we use $w$ as length scale, the P\'eclet number is $\mathrm{Pe} = 4\beta w$. Additionally, $\tau = \tau_2$ if and only if $\tau_2\geq \tau_1$; equivalently, $\tau = \tau_2$ if and only if $2\beta w \geq 1$, which entails that $\mathrm{Pe} \geq 2$.


\section{Application: A compartmental model of an infectious disease on a tie-decay network}\label{sec:sir}

Many infectious diseases spread in humans through networks of contacts between susceptible and infected people \cite{pastor2015,kiss2017}. However, not all contacts between infected and susceptible people result in an infection. If we are given the interaction pattern between individuals, we can calculate the associated tie strength of different pairs of people as a function of time. By setting a threshold on the tie strength, suppose that only contacts whose tie strength is at least as large as the threshold result in a new infection. In this section, we use a tie-decay network for the interaction patterns between individuals and simulate a compartmental model on such a network. We consider a susceptible--infected--recovered (SIR) contagion \cite{kermack1927contribution,Brauer2019}. (See \cite{qinyi2020} for analysis of susceptible--infected--susceptible (SIS) contagions on tie-decay networks.) Suppose for simplicity that individuals in a population interact with each other according to the back-to-unity tie-decay model in Section \ref{back to unity growth model}. This assumption requires that individual pair interact with equal probability, which typically does not hold in reality, and such heterogeneity can significantly affect the spread of a disease \cite{pastor2015,kiss2017}. We also assume that the disease spread does not affect the interaction pattern of the individuals. This assumption is also unrealistic, although it is likely reasonable in situations like the early stages of an epidemic. The spread of an infectious disease depends on how much and frequently individuals interact with each other. For our discussion, we also assume that when the disease first enters the population, the tie-decay network has already reached its stationary state, so we can use the distribution \eqref{eq:bounded_prob} of the tie strengths of the network at stationarity.

We follow common notation for SIR models \cite{Brauer2019}. Suppose that the population size is $N_p$. Let $S(t)$, $I(t)$, and $R(t)$ denote the (time-dependent) numbers of individuals in the susceptible, infected, and recovered compartments, respectively. The continuous-time SIR model in a well-mixed population is
\begin{align}\label{sir1}
    \frac{dS}{dt} &= -\bar{\beta} IS/N_p \,,  \nonumber \\
    \frac{dI}{dt} &= \bar{\beta}IS/N_p-\bar{\gamma}I \,, \\
    \frac{dR}{dt} &= \bar{\gamma} I \,, \nonumber
\end{align}
where $\bar{\beta}$ is the infection rate and $\bar{\gamma}$ is the recovery rate. Because it is easier to track the number of people in each compartment on a daily or weekly basis than in continuous time, it is often more meaningful to consider the following discrete version of the SIR model: 
\begin{align} \label{eq:sir_discrete}
    S_{i+1} &=S_i-\bar{\beta} I_iS_i/N_p \,, \nonumber \\
    I_{i+1} &=I_i+\bar{\beta} I_iS_i/N_p-\bar{\gamma}I_i \,, \nonumber \\
    R_{i+1} &= R_i + \bar{\gamma} I_{i}\,.
\end{align}
One interpretation of the term $\bar{\beta} I_iS_i/N_p$ is as follows. In one time step, we suppose that each susceptible individual interacts with a person who we select uniformly at random from the population. With probability $I/N_p$, this person is in the infected compartment, and such an interaction between a susceptible person and an infected person results in the former becoming infected with probability $\bar{\beta}$. The SIR models \eqref{sir1} and \eqref{eq:sir_discrete} assume that disease spread does not affect the interaction patterns of individuals. 

In the synchronous-updating SIR model \eqref{eq:sir_discrete}, each individual from the susceptible compartment interacts with one person in a single time step. By contrast, {in the SIR model on a tie-decay network}, an individual from the susceptible compartment interacts with everyone in the population with the same probability $p$. The only active interactions are ones with tie strengths that are at least as large as the threshold. The probability $\mathbf{P}$ that a tie strength is larger than the threshold is given by Eq.~\eqref{eq:bounded_prob}. In our tie-decay network setting, in one time step, an individual from the susceptible compartment can have active interactions with any other individual in the population with a probability that is given by Eq.~\eqref{eq:bounded_prob}. Therefore, the number of active interactions of a susceptible person satisfies a Poisson distribution, with mean $\lambda= N_p \mathbf{P}$, when the population $N_p$ is large and $\mathbf{P}$ small. In summary, we describe SIR disease spreading on a tie-decay network as follows. We have a population of size $N_p$. An individual in the susceptible compartment can have active interactions with each other individual in the population with a homogeneous, independent probability $\mathbf{P}$. An active interaction between a susceptible person and an infected person leads to infection with probability $\bar{\beta}$, and people from the infected compartment recover at rate $\bar{\gamma}$.

\begin{figure*}
    \begin{subfigure}[t]{0.42\textwidth}
        \centering
        \includegraphics[width=\linewidth]{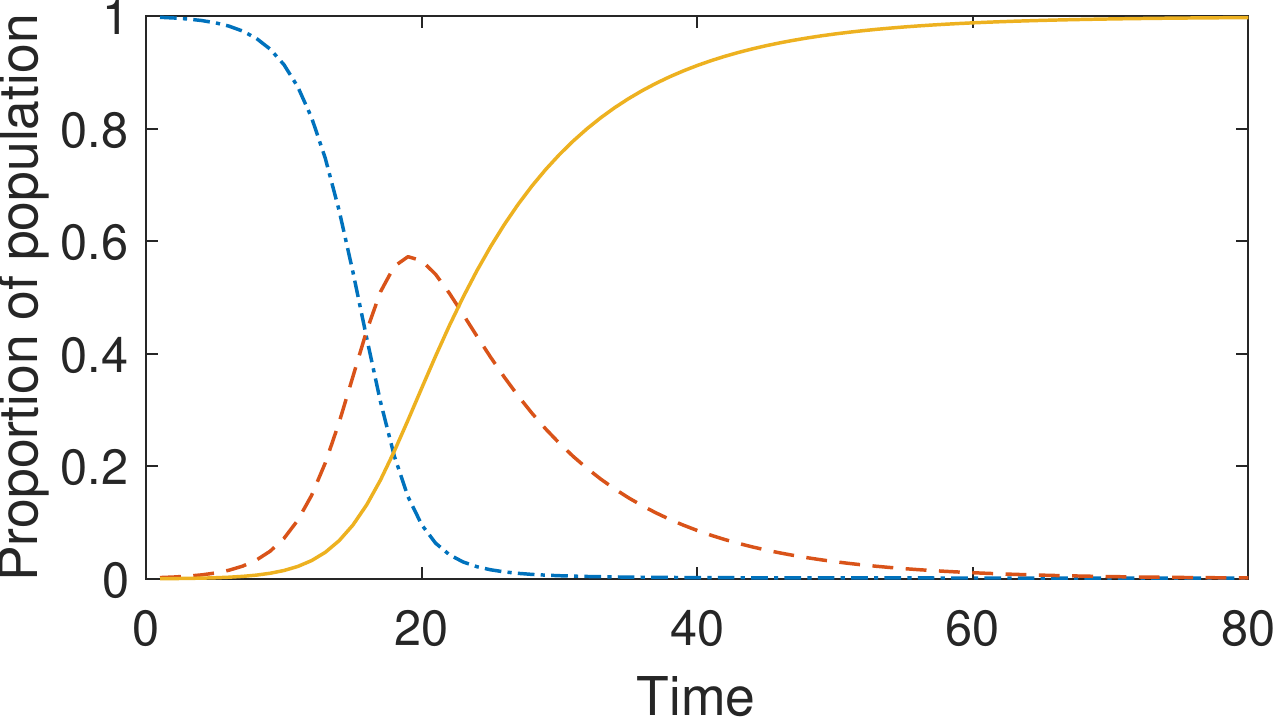}
        \caption{}
    \end{subfigure}
    \begin{subfigure}[t]{0.42\textwidth}
        \centering
        \includegraphics[width=\linewidth]{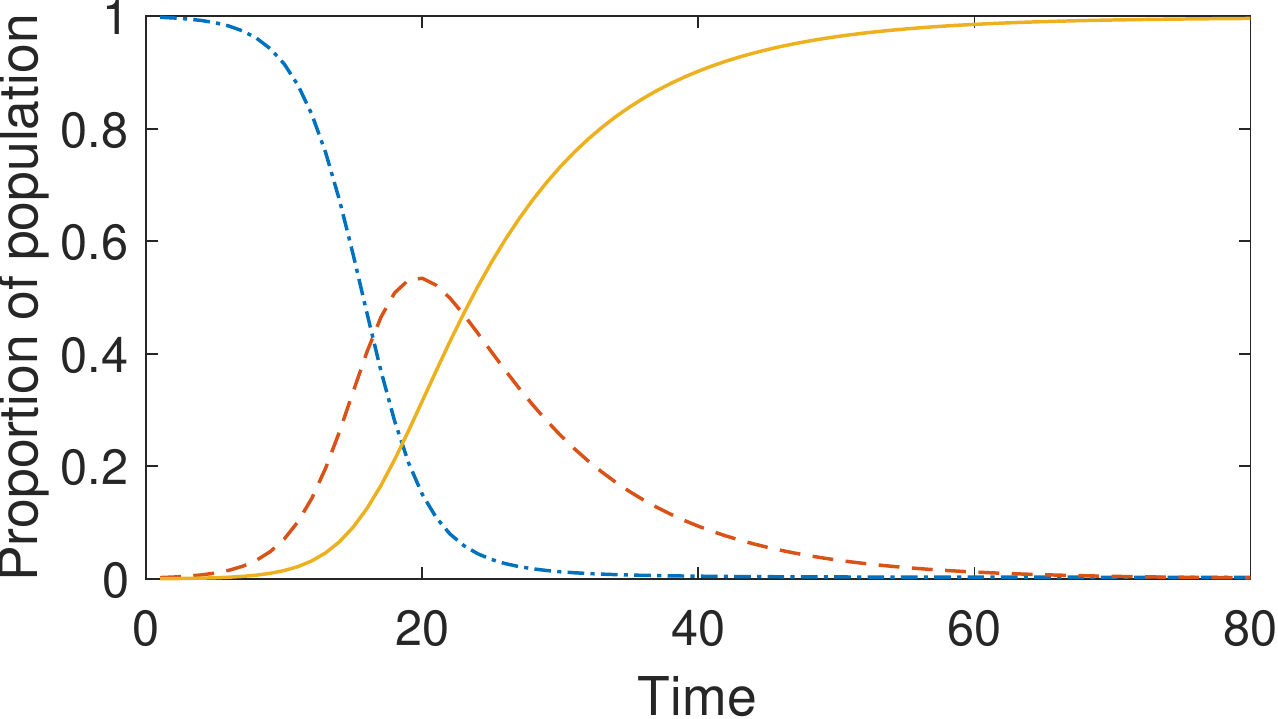}
             \caption{}
    \end{subfigure}\\
    \begin{subfigure}[t]{0.42\textwidth}
        \centering
        \includegraphics[width=\linewidth]{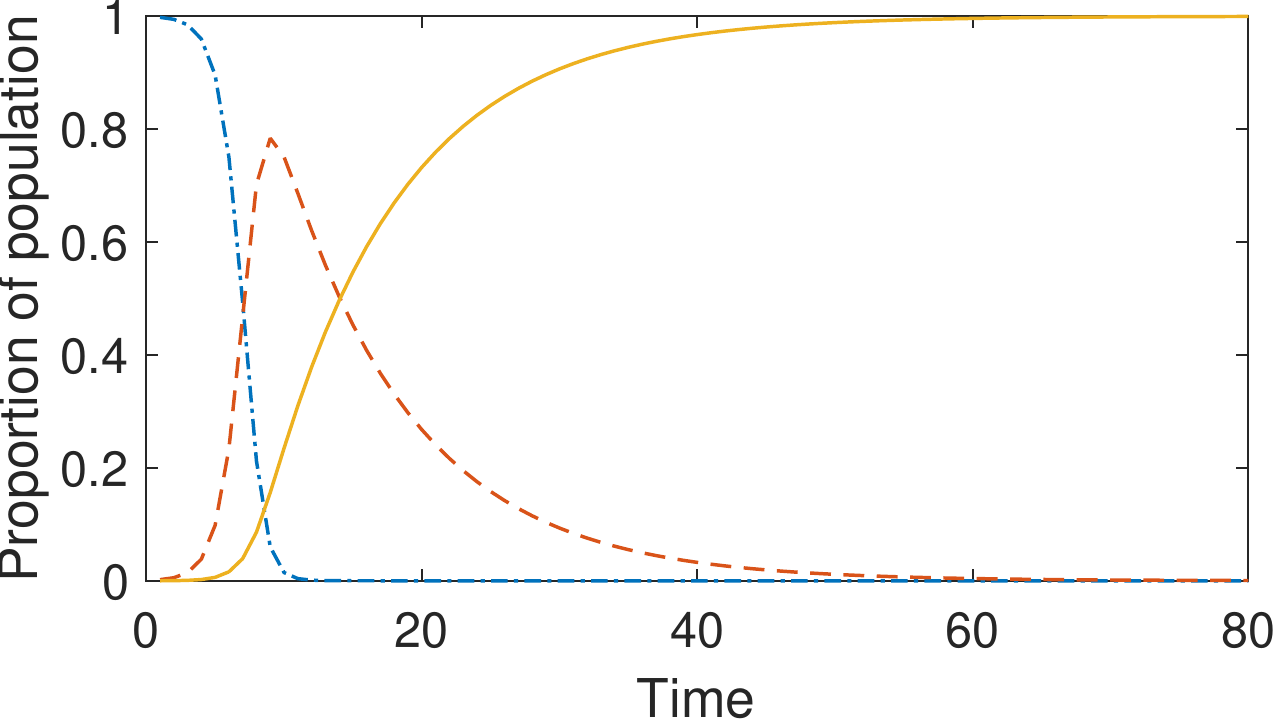}
             \caption{}
    \end{subfigure}
    \begin{subfigure}[t]{0.42\textwidth}
        \centering
        \includegraphics[width=\linewidth]{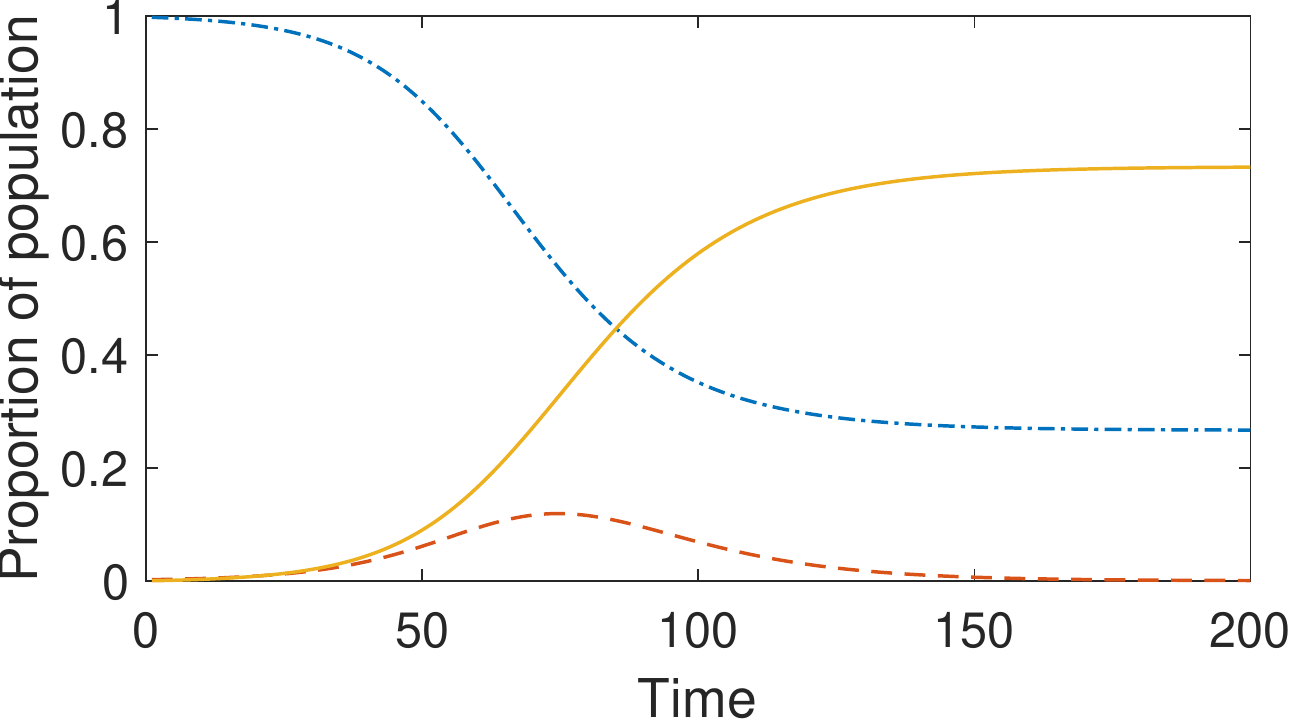}
             \caption{}
    \end{subfigure}
    \caption{Comparison of (a) the discrete SIR model \eqref{eq:sir_discrete} and (b,c,d) SIR disease spread on a tie-decay network with different values of $\lambda = N_p \mathbf{P}$, where $N_p$ is the population and $\mathbf{P}$ is the probability that a tie strength is at least the threshold. We show results for (b) $\lambda = 1$, (c) $\lambda = 3$, and (d) $\lambda = 0.3$. In each panel, the horizontal axis is time, the dot-dashed blue curve indicates susceptible individuals, the dashed red curve indicates infected individuals, and the solid yellow curve indicates recovered individuals. The infection rate is $\bar{\beta} = 0.6$, the recovery rate is $\bar{\gamma} = 0.1$, and the population is $N_p = 5000$. Our initial conditions are $S(0)/N_p = 0.998$ and $I(0)/N_p = 0.002$, and $R(0) = 0$. Observe the similarity between the plots in panels (a) and (b).
     }
    \label{fig:sir_compare}
\end{figure*}

We compare the results of simulating Eq.~\eqref{eq:sir_discrete} and SIR disease spreading on tie-decay networks for several values of $\lambda$ in Fig.~\ref{fig:sir_compare}. When $\lambda = 1$, we see that the discrete SIR model~\eqref{eq:sir_discrete} is a good approximation for SIR disease spreading on the tie-decay network. However, when $\lambda >1$, each susceptible individual interacts on average with more than one person in each time step, so the disease spreads faster on the tie-decay network than what occurs with \eqref{eq:sir_discrete}. When $\lambda <1$, a susceptible individual interacts on average with less than one person in each time step, so the disease spreads slower on the tie-decay network than what occurs with \eqref{eq:sir_discrete}. We also note that $\lambda>1$ corresponds to exactly the criterion for the existence of a GCC in the tie-decay network as $N_p \rightarrow \infty$. Therefore, we see that when there is a GCC in the tie-decay network, an SIR disease tends to spread faster than with the discrete SIR model \eqref{eq:sir_discrete}; when there is no GCC in the tie-decay network, the disease tends to spread more slowly than with the discrete SIR model. 


\section{Conclusions and discussion} 
\label{conclusions}

It is very popular to study temporal networks \cite{holme2012temporal,holme2015,Holme2019}, but most investigations of such networks focus on discrete-time approaches. However, many networks evolve continuously in time, and it is important to develop approaches for studying such temporal networks. This is an important modeling consideration, as it is often useful to consider the underlying time as continuous even when subsequently discretizing the dynamics of temporal networks.

In the present paper, we studied several continuous-time network models with tie decay, diffusion, and convection. We investigated the long-time behavior of these models and examined the emergence of giant connected components (GCCs) in the long-time limit in the networks that these models produce. In addition to exploring two existing continuous-time models --- the tie-decay model of Ahmad et al. \cite{ahmad2018tie} and a modification of the back-to-unity model of Jin et al. \cite{jin2001} --- we also developed two new models, a diffusion model and a convection--diffusion model for tie strengths, and we examined the formation of a GCC for the latter. We derived the stationary distribution of tie strengths for the convection--diffusion mechanism using intuition from numerical computations of linear convection--diffusion partial differential equations. Our analytical results agree with our numerical simulations in the long-time limit. We also examined SIR dynamics on the back-to-unity model.

All models of the models that we studied in the present paper produce temporal networks in which we distinguish between interactions and ties between nodes. In all of these models, the tie strength between two nodes grows when they interact and decays exponentially when they do not. The specific way in which tie strengths change is a key difference between the models. In the tie-decay model of \cite{ahmad2018tie}, the tie strength grows by $1$ when an interaction occurs between nodes; in the back-to-unity model of \cite{jin2001}, the tie strength grows to $1$ when there is an interaction; the convection--diffusion model, the tie strength experiences instantaneous exponential growth when there is an interaction. Another difference is the time scale of interactions between two given nodes in comparison to the time scale of a decay in their tie strength. In the Ahmad et al. tie-decay and back to unity models, it is much more likely to not have an interaction between two specified nodes in a given time step than to have one. Therefore, on average, the strength of a tie between two nodes decays for a long time after each interaction between them. By contrast, in the bounded convection--diffusion model of tie strengths from Section~\ref{bounded convection diffusion model}, the probabilities of having an interaction (namely, $\frac{1}{2}+\Delta$) and of not having an interaction (namely, $\frac{1}{2}-\Delta)$ are both very small, with the former slightly larger than the latter. Consequently, the tie strength between a given two nodes does not decay for much time before there is another interaction between them. In this model (as well as the diffusion model, which is a special case), we also impose an upper bound on tie strengths.

The tie-decay model of \cite{ahmad2018tie} and the back-to-unity model \cite{jin2001} are interesting models of interactions between people that are worth explorations in applications. It is important to consider how their interaction patterns affect dynamical processes, such as biological contagions and opinion dynamics. In Section \ref{sec:sir}, we examined susceptible--infected--recovered (SIR) dynamics on networks that are produced by the back-to-unity model. Using these types of models allows one to incorporate a variety of interaction patterns, and it is important to further study how different patterns affect dynamical processes on social networks. Current efforts include the analysis of susceptible--infected--susceptible (SIS) dynamics \cite{qinyi2020} and opinion dynamics \cite{kashin2020} on the tie-decay networks of \cite{ahmad2018tie}. Given that the tie strength between neurons can increase when they have similar interaction patterns \cite{hebbian2011}, we also expect that network models like the ones that we studied in the present paper to be relevant for analysis of phenomena like Hebbian learning in neuronal systems.

When generalizing network analysis to continuous-time formulations of temporal networks, it is important to adapt familiar network ideas to this arena. This includes random-graph models and GCCs (as in the present paper), and it will be valuable to focus future efforts on generalizing other ideas (such as community structure and dynamical processes on networks) to continuous-time network models. In our analysis, we treated edges as evolving independently, but many systems have correlations (e.g., mutual excitation or mutual inhibition) between edges, and it is important to generalize our analysis for those situations and to study how such correlations affect dynamical processes.


\section{Acknowledgements}

We thank Mariano Beguerisse D\'iaz, Heather Zinn Brooks, Tom Chou, and Hanbaek Lyu for helpful discussions. 


\clearpage


\begin{thebibliography}{27}
\providecommand{\natexlab}[1]{#1}
\providecommand{\url}[1]{\texttt{#1}}
\expandafter\ifx\csname urlstyle\endcsname\relax
  \providecommand{\doi}[1]{doi: #1}\else
  \providecommand{\doi}{doi: \begingroup \urlstyle{rm}\Url}\fi

\bibitem[Newman(2018)]{newman2018}
Mark E.~J. Newman.
\newblock \emph{Networks}.
\newblock Oxford University Press, Oxford, UK, second edition, 2018.

\bibitem[Porter(2020)]{vision2020}
Mason~A. Porter.
\newblock Nonlinearity + networks: {A} 2020 vision.
\newblock In Panayotis~G. Kevrekidis, Jes\'{u}s Cuevas-Maraver, and Avadh
  Saxena, editors, \emph{Emerging Frontiers in Nonlinear Science}, pages
  131--159. Springer International Publishing, Cham, Switzerland, 2020.

\bibitem[Holme and Saram{\"a}ki(2012)]{holme2012temporal}
Petter Holme and Jari Saram{\"a}ki.
\newblock Temporal networks.
\newblock \emph{Physics Reports}, 519\penalty0 (3):\penalty0 97--125, 2012.

\bibitem[Holme(2015)]{holme2015}
Petter Holme.
\newblock Modern temporal network theory: {A} colloquium.
\newblock \emph{European Physical Journal B}, 88:\penalty0 234, 2015.

\bibitem[Holme and Saram{\"a}ki(2019)]{Holme2019}
Petter Holme and Jari Saram{\"a}ki.
\newblock \emph{{Temporal Network Theory}}.
\newblock Springer International Publishing, {Cham, Switzerland}, 2019.

\bibitem[Ahmad et~al.(2018)Ahmad, Porter, and
  Beguerisse-D{\'\i}az]{ahmad2018tie}
Walid Ahmad, Mason~A. Porter, and Mariano Beguerisse-D{\'\i}az.
\newblock Tie-decay temporal networks in continuous time and eigenvector-based
  centralities.
\newblock \emph{arXiv preprint arXiv:1805.00193}, 2018.

\bibitem[Burt(2000)]{burt2000decay}
Ronald~S. Burt.
\newblock Decay functions.
\newblock \emph{Social Networks}, 22\penalty0 (1):\penalty0 1--28, 2000.

\bibitem[Lerman(2016)]{lerman2016information}
Kristina Lerman.
\newblock Information is not a virus, and other consequences of human cognitive
  limits.
\newblock \emph{Future Internet}, 8\penalty0 (2):\penalty0 21, 2016.

\bibitem[Lerman et~al.(2012)Lerman, Ghosh, and Surachawala]{lerman2012social}
Kristina Lerman, Rumi Ghosh, and Tawan Surachawala.
\newblock Social contagion: An empirical study of information spread on digg
  and twitter follower graphs.
\newblock \emph{arXiv preprint arXiv:1202.3162}, 2012.

\bibitem[Navarro et~al.(2017)Navarro, Miritello, Canales, and Moro]{moro2017}
Henry Navarro, Giovanna Miritello, Arturo Canales, and Esteban Moro.
\newblock Temporal patterns behind the strength of persistent ties.
\newblock \emph{European Physical Journal --- Data Science}, 6\penalty0
  (1):\penalty0 31, 2017.

\bibitem[Agliari and Barra(2011)]{hebbian2011}
E.~Agliari and A.~Barra.
\newblock A {H}ebbian approach to complex-network generation.
\newblock \emph{Europhysics Letters}, 94\penalty0 (1):\penalty0 10002, 2011.

\bibitem[Jin et~al.(2001)Jin, Girvan, and Newman]{jin2001}
Emily~M. Jin, Michelle Girvan, and Mark E.~J. Newman.
\newblock Structure of growing social networks.
\newblock \emph{Physical Review E}, 64:\penalty0 046132, 2001.

\bibitem[Fritz et~al.(2019)Fritz, Lebacher, and Kauermann]{fritz2019}
Cornelius Fritz, Michael Lebacher, and G\"{o}ran Kauermann.
\newblock Tempus volat, hora fugit --- {A} survey of dynamic network models in
  discrete and continuous time.
\newblock \emph{arXiv preprint arXiv:1905.10351}, 2019.

\bibitem[Sulo et~al.(2010)Sulo, Berger-Wolf, and Grossman]{sulo2010meaningful}
Rajmonda Sulo, Tanya Berger-Wolf, and Robert Grossman.
\newblock Meaningful selection of temporal resolution for dynamic networks.
\newblock In \emph{Proceedings of the Eighth Workshop on Mining and Learning
  with Graphs}, pages 127--136. ACM, 2010.

\bibitem[Erd\H{o}s and R{\'e}nyi(1960)]{erdosrenyi1960evolution}
Paul Erd\H{o}s and Alfr{\'e}d R{\'e}nyi.
\newblock On the evolution of random graphs.
\newblock \emph{Publication of the Mathematical Institute of the Hungarian
  Academy of Sciences}, 5\penalty0 (1):\penalty0 17--60, 1960.

\bibitem[Saberi(2015)]{saberi2015}
Abbas~Ali Saberi.
\newblock Recent advances in percolation theory and its applications.
\newblock \emph{Physics Reports}, 578:\penalty0 1--32, 2015.

\bibitem[L\'opez et~al.(2007)L\'opez, Parshani, Cohen, Carmi, and
  Havlin]{lopez2007}
Eduardo L\'opez, Roni Parshani, Reuven Cohen, Shai Carmi, and Shlomo Havlin.
\newblock Limited path percolation in complex networks.
\newblock \emph{Physical Review Letters}, 99:\penalty0 188701, 2007.

\bibitem[Pastor-Satorras et~al.(2015)Pastor-Satorras, Castellano, Van~Mieghem,
  and Vespignani]{pastor2015}
Romualdo Pastor-Satorras, Claudio Castellano, Piet Van~Mieghem, and Alessandro
  Vespignani.
\newblock Epidemic processes in complex networks.
\newblock \emph{Reviews of Modern Physics}, 87:\penalty0 925--979, 2015.

\bibitem[Gleich(2015)]{gleich2015}
D.~F. Gleich.
\newblock {PageRank beyond the Web}.
\newblock \emph{{SIAM Review}}, 57\penalty0 (3):\penalty0 321--363, 2015.

\bibitem[Evans(2010)]{evans1997partial}
Lawrence~C. Evans.
\newblock \emph{Partial Differential Equations}, volume~19 of \emph{Graduate
  Studies in Mathematics}.
\newblock American Mathematical Society, Providence, RI, USA, 2010.

\bibitem[Levin and Peres(2017)]{levin2017markov}
David~A. Levin and Yuval Peres.
\newblock \emph{Markov Chains and Mixing Times}.
\newblock American Mathematical Society, Providence, RI, USA, 2017.

\bibitem[Batchelor(1967)]{batchelor1967introduction}
George~K. Batchelor.
\newblock \emph{An Introduction to Fluid Dynamics}.
\newblock Cambridge University Press, Cambridge, UK, 1967.

\bibitem[Kiss et~al.(2017)Kiss, Miller, and Simon]{kiss2017}
Istv{\'a}n~Z. Kiss, Joel~C. Miller, and P{\'e}ter~L. Simon.
\newblock \emph{{Mathematics of Epidemics on Networks: From Exact to
  Approximate Models}}.
\newblock Springer International Publishing, {Cham, Switzerland}, 2017.

\bibitem[Kermack and McKendrick(1927)]{kermack1927contribution}
William~Ogilvy Kermack and Anderson~G McKendrick.
\newblock A contribution to the mathematical theory of epidemics.
\newblock \emph{Proceedings of the Royal Society of London. Series A},
  115\penalty0 (772):\penalty0 700--721, 1927.

\bibitem[Brauer et~al.(2019)Brauer, Castillo-Chavez, and Feng]{Brauer2019}
Fred Brauer, Carlos Castillo-Chavez, and Zhilan Feng.
\newblock \emph{Mathematical Models in Epidemiology}.
\newblock Springer-Verlag, Heidelberg, Germany, 2019.

\bibitem[Chen and Porter()]{qinyi2020}
Qinyi Chen and Mason~A. Porter.
\newblock Epidemic thresholds of infectious diseases on tie-decay networks.
\newblock In preparation.

\bibitem[Sugishita et~al.()Sugishita, Porter, Beguerisse-D\'{i}az, and
  Masuda]{kashin2020}
Kashin Sugishita, Mason~A. Porter, Mariano Beguerisse-D\'{i}az, and Naoki
  Masuda.
\newblock Opinion dynamics in tie-decay networks.
\newblock In preparation.

\end{thebibliography}



\end{document}